\documentclass[a4paper,10pt,nofootinbib]{article}
\usepackage[margin=2.84cm]{geometry}

\usepackage{mathptmx}
\usepackage{amsmath,amssymb}
\usepackage[normalem]{ulem}
\usepackage{graphicx}
\usepackage[svgnames]{xcolor}
\usepackage[sort&compress,numbers]{natbib}

\usepackage[colorlinks=true,urlcolor=MediumBlue,linkcolor=Crimson,citecolor=Crimson]{hyperref}
\usepackage{cleveref}
\usepackage{url}
\usepackage{feynmp}
\usepackage{float}
\usepackage{subcaption}
\usepackage{booktabs}
\usepackage{longtable}

\newcommand{\freq}{f}

\newcommand{\beginsupplement}{%
        \setcounter{table}{0}
        \renewcommand{\thetable}{S\arabic{table}}%
        \setcounter{figure}{0}
        \renewcommand{\thefigure}{S\arabic{figure}}%
     }
\title{\bf Axion Dark Matter: What is it and Why Now?}
\author{\bf Francesca Chadha-Day$^{1, \ast}$, John Ellis$^{2,3, \ast}$, David J. E. Marsh$^{2, \ast}$}

\begin{document}
\maketitle

\begin{abstract}
The axion has emerged in recent years as a leading particle candidate to provide the mysterious dark matter in the cosmos, as we review here for a general scientific audience. We describe first the historical roots of the axion in the Standard Model of particle physics and the problem of charge-parity invariance of the strong nuclear force. We then discuss how the axion emerges as a dark matter candidate, and how it is produced in the early Universe. The symmetry properties of the axion dictate the form of its interactions with ordinary matter. Astrophysical considerations restrict the particle mass and interaction strengths to a limited range, which facilitates the planning of experiments to detect {the axion}. A companion review discusses the exciting prospect that the axion could {indeed} be detected in the near term in the laboratory.\\
   ~~\\
   {\it A review prepared for Science Advances}\\
   ~~\\
$^1$ Institute for Particle Physics Phenomenology, Department of Physics, Durham University, \\ Durham DH1~3LE, United Kingdom\\
$^2$ Theoretical Physics and Cosmology Group, Department of Physics, King’s College London, Strand, \\ London WC2R 2LS, United Kingdom\\
$^3$ Theoretical Physics Department, CERN, CH-1211 Geneva 23, Switzerland; \\
National Institute of Chemical Physics \& Biophysics, R{\"a}vala 10, 10143 Tallinn, Estonia\\
$^{\ast}$ To whom correspondance should be addressed \\
~~\\
KCL-PH-TH/2021-20, CERN-TH-2021-045, IPPP/20/91
\end{abstract}

\section{Introduction}

The Standard Model of particle physics provides a successful description of the visible matter in the Universe, from stars to the inner workings of atoms and nuclei.
It categorises the fundamental constituents of matter, the quarks and leptons, and the electromagnetic, weak and strong nuclear forces between them.
The Standard Model is a mathematically consistent quantum field theory that allows theorists to calculate accurate predictions, which have in many cases been
verified experimentally with a precision below the per-mille level at the Large Hadron Collider (LHC) and other particle accelerators. 
The crowning success of the Standard Model was the discovery of the Higgs boson in 2012~\cite{ATLASH,CMSH}, a particle of a novel type
whose existence was predicted in 1964~\cite{Higgs:1964pj} in order to solve theoretical problems associated with the masses
of vector bosons, see also~\cite{Englert:1964et}. Precise experimental measurements in the decades preceding 2010
verified many other predictions of the Standard Model, including the existence and mass of the top quark.
They also confirmed the necessity of the Higgs boson and enabled its mass to
be estimated numerically. Following its discovery, experiments have confirmed that it {has zero spin}, unlike any
fundamental particle {known previously}, {and interactions with} other particles that are proportional to their masses, as predicted by the
Standard Model.

Despite these manifold successes,
particle physicists are restless and dissatisfied with the Standard Model, because it has many theoretical shortcomings and leaves many pressing experimental
questions unanswered. Here we review the axion~\cite{pecceiquinn1977,weinberg1978,wilczek1978}, a hypothetical particle beyond the Standard Model that addresses some of these unresolved issues and
is the focus of growing experimental interest.

One of these issues is how matter and antimatter are (and are not) distinguished at the level of fundamental particles. For each matter particle in the Standard Model, special relativity and quantum mechanics require~\cite{Dirac:1928hu} the existence of a corresponding anti-matter particle with identical mass and spin but opposite charge. {Uncharged particles such as photons may be their own anti-particles.} The existence of every anti-particle in the Standard Model {has been} confirmed experimentally by {observing their production in the collisions of} ordinary particles. However, astrophysical and cosmological
observations tell us that most of the visible material in the Universe is composed of the same matter particles as us on Earth, and that there are no
large concentrations of antimatter. This cosmological matter-antimatter asymmetry is thought to be due to differences in the interactions of
elementary particles and antiparticles~\cite{Sakharov:1967dj}, which violate certain symmetries that distinguish particle from antiparticle. These are charge conjugation (denoted by $C$) and its combination with parity reversal (denoted by $P$, also known as spatial inversion symmetry).
A form of $CP$ violation was discovered in the laboratory over 50 years ago~\cite{Christenson:1964fg}, and by now it has been observed in many {decays via} the weak interactions.
These observations can be accommodated within the Standard Model with six quarks~\cite{Kobayashi:1973fv}, albeit without a profound explanation. However, this Standard Model 
mechanism is insufficient to explain the cosmological matter-antimatter asymmetry and, moreover, it is a puzzle that $CP$ violation does not appear also
in the strong nuclear interaction: this is known as the ``strong-$CP$ problem''. 
As we discuss below, {\em this puzzle could be resolved via the hypothetical axion particle}. In this review, we reserve the term `axion' for the particle arising from the solution to the strong-$CP$ problem described below, also known as the `QCD axion'. This will be the main topic of this review. We will also touch briefly on `axion-like particles', which do not solve the strong-$CP$ problem. As a result, the properties of axion-like particles are less theoretically constrained.

The failure to explain the cosmological matter-antimatter asymmetry and the $CP$ symmetry of the strong force are just two of the Standard Model's failures to describe the Universe. {Some might regard them} as ``cosmetic'' problems, since they could be solved by fiat within the context of the Standard Model.  There is, however, a much more pressing existential problem.

Multiple astronomical observations tell us that the visible matter described by the Standard Model provides just a small part of the total
density of the Universe. There is much more invisible ``dark matter'' out there~\cite{Rubin:1970zza} that remains to be explained. This is {gross inadequacy of} the Standard Model, since {the dark matter cannot be explained} without appealing to substantial amounts of new physics. The existence of dark matter is inferred from its
gravitational effects, and astrophysical observations suggest that it is ``cold", i.e., it has been moving very slowly for much of the history
of the Universe, and there are experimental upper limits on how strongly it interacts with the visible matter. Moreover, the agreement between independent cosmological observations (including galaxy clustering and the temperature anisotropies in the cosmic microwave background, CMB, radiation~\cite{Aghanim:2018eyx}) and the corresponding theory of structure formation also tells us that dark matter must have been present since early in the history of the Universe, 
a year or so after the Big Bang or even earlier. {Jim Peebles' share of the 2019 Nobel Prize in Physics was for the theory of cosmological structure formation behind this understanding, and the evidence it provides for the existence of cold dark matter~\cite{Peebles:1982ff}.}
Beyond these basic facts relating to the temperature and longevity of the dark matter, we have very little information about its nature and properties. 
It may well consist of one or more types of fundamental particle, though part or
all of it might consist of macroscopic lumps of some invisible form of matter such as black holes. {\em The axion has become one of the prime particle
candidates for providing dark matter}.

Historically, the first type of particle dark matter that was suggested was some sort of weakly-interacting massive particle, abbreviated to WIMP,
such as a massive neutrino. Minimal versions of this hypothesis within the Standard Model were excluded experimentally, leading particle theorists
to consider possible candidates in extensions of the Standard Model, e.g., based on supersymmetry~\cite{Jungman:1995df}. However, this is not the only possibility.
If the dark matter particle is a low-mass boson, it may populate the Universe in a coherent wave-like state that is also slow-moving and hence ``cold". 
{\em The axion is the prototype for such wave-like dark matter}.

\begin{figure}
    \centering
    \includegraphics[width=\textwidth]{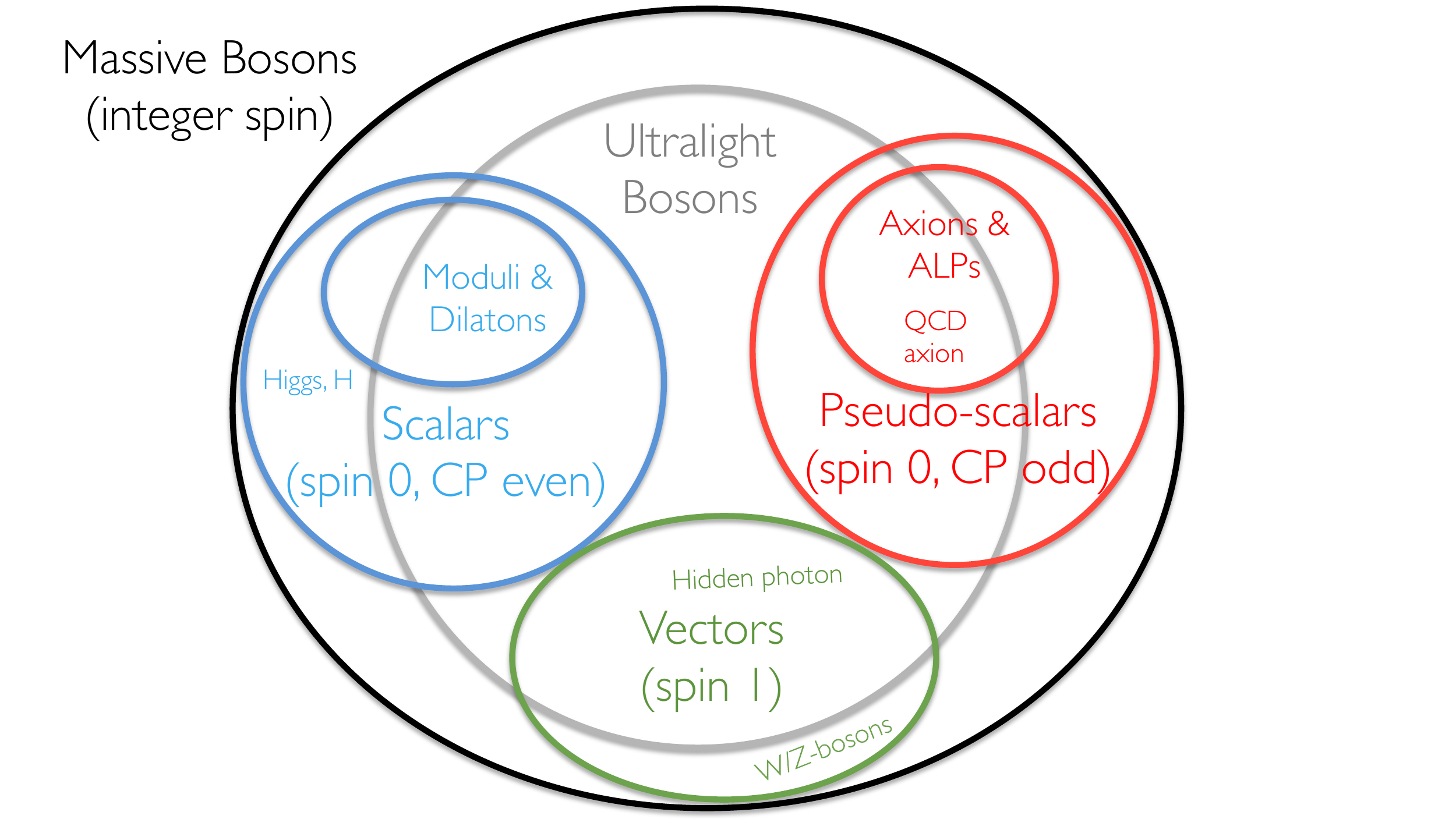}
    \caption{Many extensions of the Standard Model predict additional massive bosons, beyond the $W, Z$ and Higgs
    bosons of the Standard Model. They might be scalar (even under both P and CP transformations), pseudoscalar (odd under both P and CP)
    or vector particles. The prototype for a
    pseudoscalar boson is the axion, which is ``ultralight'' with mass $m\ll 1\text{ eV}$. Other proposals include pseudoscalar axion-like particles (ALPs), hidden photons, and
    scalar particles such as moduli and dilatons. Any of these might provide the astrophysical dark matter.}
    \label{fig:venn}
\end{figure}

For many years following the proposals of the axion and WIMP in the late 1970s, axion dark matter was a minority interest among particle
physicists, who were mainly focused on high-energy collider physics. Their primary motivation was the search for the Higgs boson, but WIMP searches
were also an important consideration that could often be pursued in parallel with the Higgs search. The search for the Higgs boson culminated in its discovery at the LHC in 2012, but WIMP searches at the LHC
and elsewhere~\cite{Aprile:2018dbl} have proved fruitless so far. These two developments have contributed to growing theoretical and experimental interest in the axion.
On the one hand, the Higgs is an existence proof for an apparently elementary spin-zero boson undergoing spontaneous symmetry breaking, something that was somewhat controversial and
unknown previously. This proof of principle invalidated one possible theoretical prejudice against the existence of the axion, which shares these properties with the Higgs.
Perhaps more importantly, the non-appearance of any WIMP at the LHC and in other direct searches has diminished enthusiasm for that candidate for particle dark matter, though
some hope still springs eternal. In parallel, there has been growing realization that axion-like particles (ALPs) appear quite generically in
extensions of the Standard Model, e.g., those with their roots in string theory, see Fig.~\ref{fig:venn}. For these reasons, {\em the axion has now become a favoured
theoretical candidate for dark matter}, motivating this review. 

A companion article reviews the
prospects for axion and ALP searches driven by experimental and technological advances at the so-called ``precision frontier'' of particle physics. These advances make axion {dark matter} a viable candidate for discovery, which until recently appeared impossible, and {are important factors} driving {the growing} interest in the field.

\fbox{
\vspace{10mm}
\begin{minipage}{5in}
\textbf{What is known about dark matter {(DM)}?} 
\begin{itemize}
    \item Cosmic density ({strong} evidence: cosmic microwave background anisotropies~\cite{Aghanim:2018eyx}). Expressed as a fraction of the {total} density of the Universe, DM makes up 26\% of the Universe, compared to 6\% in ordinary matter, and 68\% in vacuum energy.
    \item Local density ({strong} evidence: Milky Way stellar motions). The local density of dark matter is around 0.3 to 0.4~GeV~cm$^{-3}$, equivalent to one proton every few cubic centimetres, or one solar mass per cubic lightyear. The density is measured on average over a relatively large fraction of the galaxy. The actual density at the precise location of the Earth could be significantly different. This is particularly relevant to axions, as discussed below. The local density is around $10^5$ times the average cosmic density.
    \item Local velocity dispersion ({strong} evidence: Milky Way stellar motions). The velocity dispersion of DM is around $\sigma_v=200\text{ km s}^{-1}$, and our local motion with respect to the galactic rest frame is in the direction of the constellation Cygnus.
    \item No preferred galactic length scale (strong evidence: galaxy clustering and evolution). DM must be non-relativistic ($v\sim c$ would allow DM to move significant distances during galaxy formation), and have negligible pressure (which would imprint sound waves during galaxy formation). This discounts standard model neutrinos and other ``hot'' or ``warm'' DM. For bosons, the de Broglie wavelength (which can be modelled as an effective pressure) must be small compared to the galaxy clustering scale.
    \item Early appearance of DM (strong evidence: galaxy clustering). DM had to be present, and gravitating, in the Universe long before the cosmic microwave background formed, and its gravitational influence began before the Universe was one year old. For light bosonic DM (like the axion) this corresponds to the latest epoch of particle creation ($t_\textup{cold}$ in Fig.~\ref{fig:two_roads}).
    \item Lack of significant interactions (strong evidence: the ``Bullet Cluster''~\cite{Clowe:2003tk}). DM cannot interact with itself or ordinary matter too strongly.
    
\end{itemize}

\end{minipage}
}
\vspace{5mm}\\

\section{Enter the Axion}

\subsection{The Strong $CP$ Problem}

The strong $CP$ problem {may be posed simply as follows:} why have we observed no electric dipole moment (EDM) for the neutron? The neutron EDM is a quantity that would violate Charge Parity ($CP$) symmetry. This is in contrast to most of the laws of fundamental physics, which look the same after a $CP$ transformation. 

To understand the effect of a $CP$ transformation on the neutron, we can instead think about how a neutron would behave under a time reversal transformation. This is because if quantum field theory is Lorentz invariant (i.e., if it is consistent with Einstein's theory of Special Relativity), then all particles and processes {\it must} be invariant under the $CPT$ combination of discrete transformations,
where $T$ reverses the arrow of time (this is known as the $CPT$ theorem of quantum field theory). So, if we find something that violates $T$ symmetry, it must also violate $CP$ symmetry in such a way that the combination $CPT$ is {\it not} violated. 

As an illustration, let us consider the effect of time reversal on the energy of a neutron with an electric dipole moment ${\bf d}$ in an electric field {\bf E}. This energy is given by the Hamiltonian

\begin{equation}
    \mathcal{H} = -{\bf d} \cdot {\bf E}.
    \label{eqn:EDMenergy}
\end{equation}
If the neutron has a permanent EDM, this EDM must be aligned with the neutron's spin. {(This follows from the Wigner-Eckart theorem in quantum mechanics.)} When we reverse the direction of time, the direction of the neutron's spin is reversed, and therefore the direction of the EDM must also be reversed under a time reversal transformation $T$. However, the electric field direction remains the same under a time reversal transformation. So we can see from equation \eqref{eqn:EDMenergy} that if the neutron has a non-zero permanent EDM, then the energy of a neutron in an electric field will change under a $T$ transformation. {Hence} a neutron electric dipole moment would violate $T$ symmetry and therefore also $CP$ symmetry. 

\begin{figure}
    \centering
    \includegraphics[width=0.95\textwidth]{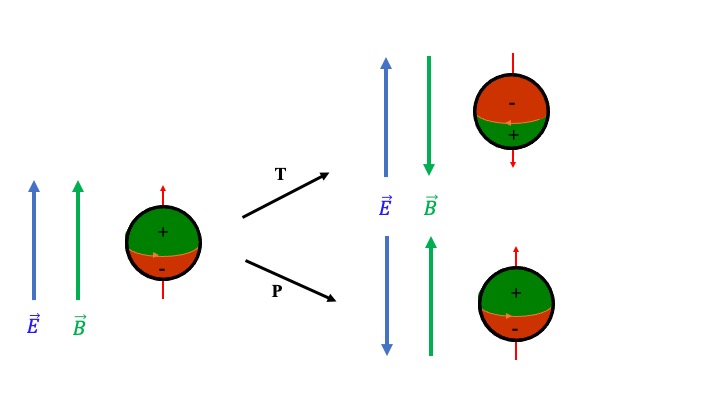}
    \caption{Effect of time and parity transformations an electric or magnetic dipole moment (proportional to the spin shown by the red arrow), on an electric field and on a magnetic field.}
    \label{fig:EDM}
\end{figure}

Figure \ref{fig:EDM} shows the $P$ and $T$ transformations of a permanent electric or magnetic dipole moment (which must be aligned with a particle's spin), of an electric field and of a magnetic field. We see that an electric dipole moment violates both $T$ and $P$ symmetry, as the spin transforms with the opposite sign to the electric field under both these transformations. Conversely, a {\it magnetic} dipole moment does not violate $P$ or $T$ symmetry as the spin transforms in the same way as the magnetic field.

Some experimentally observed processes within the Standard Model {\it do} violate $CP$ symmetry, i.e., the process and the corresponding CP transformed process occur at different rates. These $CP$-violating processes are generated by the weak interaction. However, $CP$ violation has {\it not} been observed in any experiment on the strong interaction, which describes the forces that bind quarks together to form protons, neutrons and other hadrons in the theory of Quantum Chromodynamics (QCD). This is very surprising because the Standard Model predicts that the structure of the QCD vacuum itself should violate $CP$. The $CP$ violation of the QCD vacuum can be parameterized by an angle denoted $\theta$.

Furthermore, a neutron EDM should {\it also} receive a contribution from the effect of the weak interaction (the interaction responsible for radioactive decays) on the quarks. Thus the total neutron EDM can be expressed using the total $CP$-violating angle $\Bar{\theta}$:

\begin{equation}
    \Bar{\theta} = {\rm strong} \, {\rm interaction
    } \, {\rm contribution} + {\rm weak} \, {\rm interaction} \, {\rm contribution}
\end{equation}
 Within the Standard Model, the electric dipole moment of the neutron is proportional to $\Bar{\theta}$, with $|{\bf d}| \sim 3.6\times 10^{-16} \Bar{\theta}\, e\, {\rm cm}$. The two contributions making up $\Bar{\theta}$ are entirely unrelated within the Standard Model and are simply set by hand. Note that, although $\Bar{\theta}$ is the sum of these two parameters, only the {\it total} $\Bar{\theta}$ is experimentally observable. There is no way to measure the strong interaction and weak interaction contributions to $\Bar{\theta}$ individually. Experiment tells us that $|{\bf d}| < 1.8 \times 10^{-26}  e\, {\rm cm}$~\cite{Abel:2020gbr}, implying that $\Bar{\theta} < 5 \times 10^{-11}$. Thus, a very precise cancellation between unrelated parameters is required to explain the observations. This could merely be a coincidence, or it could be a first hint of new fundamental physics.

\subsection{The Axion Solution}

\fbox{
\vspace{10mm}
\begin{minipage}{5in}
\textbf{Strong $CP$ Problem and the Axion} 
\begin{itemize}
    \item The Standard Model contains a constant, $\bar{\theta}$, which is known to be an angle and thus take values between 0 and $2\pi$. Symmetry specifies no preferred angle.
    \item The value of this angle determines the neutron electric dipole moment, $|{\bf d}| = 3.6\times 10^{-16}\bar{\theta}\, e\, {\rm cm}$. Measurements are consistent with $\bar{\theta}\lesssim 10^{-10}$, which suggests some missing symmetry principle.
    \item The axion theory introduces a new field, $a(t,x)$, such that $\bar{\theta}\propto a(t,x)$, and for which the value $a=0$ is energetically favourable.
\end{itemize}

\end{minipage}
}
\vspace{5mm}\\

The axion is the most popular solution to the strong $CP$ problem. It {appeared first in a model} proposed by Roberto Peccei and Helen Quinn in 1977~\cite{pecceiquinn1977}. Their idea was to promote $\Bar{\theta}$ to a dynamical quantity, rather than simply a constant parameter {as in} the Standard Model. In practice, this means adding a new field - the axion field -  to the Standard Model that couples to the strong nuclear force in the same way as $\Bar{\theta}$. As we shall see, the axion field relaxes to a value such that $\Bar \theta$ is zero and hence the total neutron EDM vanishes.

The behaviour of particles is determined by their symmetries. Emmy Noether showed that these symmetries correspond to conservation laws, such as conservation of energy or charge. In the Standard Model, symmetries lead to forces between particles: the force ``communicates'' the symmetry from place to place between particles. Forces are in turn mediated by bosonic particles: the force is a field, and quantum mechanics associates a particle with every field (e.g. the photon for the electromagnetic field and electromagnetic gauge symmetry). Thus, to add the axion to the Standard Model, we introduce a new symmetry, called the Peccei-Quinn (PQ) symmetry~\cite{pecceiquinn1977}, which is a {\it global axial $U(1)$} symmetry. `Global' means that the symmetry transformation is the same everywhere and for all time - in contrast to the gauge symmetries associated with interactions in the Standard Model, whose transformations can be made independently at different points and at different times. `Axial' means that the symmetry transformation acts differently on left-handed and right-handed particles. Roughly speaking, a particle's handedness tells us about the relative orientation of its spin and its velocity. Finally, a $U(1)$ symmetry transformation is one that is mathematically equivalent to a rotation about a single axis.  \\

Introducing the PQ symmetry corresponds to introducing a new bosonic field beyond the single Higgs boson of the Standard Model. At high temperatures, such as in the early universe, the PQ symmetry is evident in all particle interactions. However, there is a phase transition that ``hides'' the PQ symmetry when the temperature falls low enough (the bosonic force carriers of the symmetry become heavy and cannot be thermally excited). Similar phenomena of ``spontaneous symmetry breaking" are central to our understanding of many areas of macroscopic physics, such as superconductivity, and there are also examples in particle physics, such as the Higgs
mechanism for generating particle masses.

\begin{figure}
    \centering
    \includegraphics[width=0.9\textwidth]{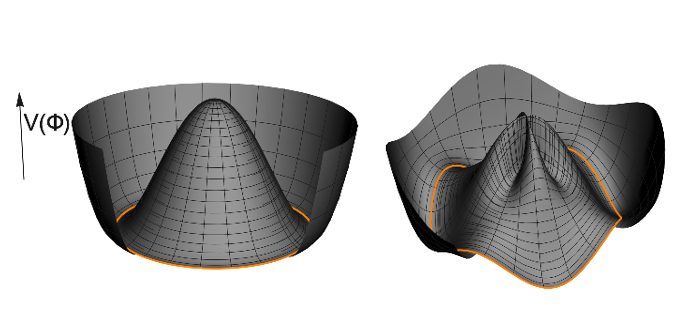}
    \caption{``Sombrero" potential of the Peccei-Quinn field $\Phi$ {is shown schematically} before (left) and after (right) the QCD phase transition. The axion corresponds to the angular direction of this potential, shown by the orange line. The state of the field is given by a point in the potential. Low energy configurations are favoured. For illustration, the potential on the right is shown for a scenario with a large amount of PQ symmetry breaking. More details are given in appendix~\ref{sec:InvisibleAxionModels}.}
    \label{fig:potential}
\end{figure}

To visualize spontaneous symmetry breaking, consider a ball rolling in the ``sombrero" potential shown on the left of Fig.~\ref{fig:potential}. If the ball has enough energy (as at high temperature), it is able to roll over the hill in the centre of the potential and occupies equally all areas of the circular valley in the ``brim" of the potential.
Now imagine that the ball loses energy (as at low temperature), slows down and comes to rest. It will choose at random  to sit in one {\it particular} position in the potential well - even though every position in the circular minimum of the potential well is exactly equivalent. This arbitrary choice (a ``choice'' made the random thermal state of the early Universe) is spontaneous symmetry breaking. Notice now that, while it would take a lot of energy to get the ball over the potential hill again, we can push the ball {\it around} the circle of the potential well with the smallest of nudges. This is a generic feature of spontaneous symmetry breaking. In particle physics, it corresponds to the appearance following spontaneous symmetry breaking
of a {\it massless} particle, which is called a Nambu-Goldstone boson~\cite{Nambu:1960xd,Goldstone:1961eq}. The Nambu-Goldstone boson of the spontaneously broken Peccei-Quinn symmetry is the axion~\cite{weinberg1978,wilczek1978}. It is represented by a field, $a$, which is proportional to the problematic $\Bar{\theta}$ angle of the strong-$CP$ problem, making the angle dynamical rather than a fixed and mysterious constant. (Achieving this remarkable theoretical sleight of hand to make $a\propto \Bar{\theta}$ is described briefly in what follows, and in more detail in the Supplemental Material.)\\

We have not yet solved the Strong $CP$ problem, as the massless axion field could {\it a priori} take any value. The next part of the story is the QCD phase transition (strictly speaking, a cross over) that occurs as the temperature falls. When the temperature is sufficiently low, QCD becomes strongly-coupled and confines quarks and gluons into the bound-state protons, neutrons and other hadrons that we see today. This phase transition breaks the PQ symmetry by distorting the sombrero potential as seen on the right in Fig.~\ref{fig:potential}. The potential now has discrete minima and the energy is minimised by the axion field taking the value of one of these minima. Thus, after the QCD cross over, the axion field rolls to the newly created minimum point, which is where the contribution of $\Bar{\theta}$ to the neutron EDM vanishes, setting the net neutron EDM to zero. (The reason this minimum has the right $CP$ properties is discussed in the Supplemental Material.) Notice that now, to make the ball move around the sombrero potential we would need to push it away and up from its minimum point. The energy required to move the classical axion field a small distance away from the minimum can be modelled as an effective potential $V(a)=m_a^2 a^2/2$. Upon quantisation, we interpret the parameter $m_a$ in the classical potential as the mass of the axion particle. 

The axion mass can be computed in terms of well-understood physics of the strong nuclear force by considering the axion mixing with the neutral pion - a bound state of quarks with the same quantum numbers as the axion. The axion's interactions with the neutral pion mean that the pion's mass generates a small mass for the axion - this effect is only possible because the axion and the pion have the same quantum numbers. This leads to the following relation for the axion mass $m_a$:
\begin{equation}
m_a f_a \sim m_{\pi} f_{\pi},
\label{eq: mass}
\end{equation}
where $f_a$ is proportional to the energy at which the PQ symmetry is spontaneously broken, $m_{\pi}$ is the pion mass, and $f_{\pi}$ is a constant
that is known from the rate of decay of the pion via the weak interaction. (A deeper understanding of the origin of the axion mass relies on the theory of ``instantons'', 
strong-interaction effects that are discussed briefly in the Supplemental Material.) Using the experimental measurements of these pion properties we can calculate the axion mass. {The axion mass, as most scales in particle physics, is typically given in electronvolts (eV), where $1\text{ eV}\approx 1.8\times 10^{-36}\text{ kg}$ which assumes $\hslash=c=1$, but we restore units of $\hslash$ and $c$ in formulae for clarity. Another useful conversion is between mass in electronvolts and Compton frequency, such that $0.25\text{ Hz}\approx 10^{-15}\text{ eV}$.} The axion mass is: 
\begin{equation}
    m_a = (5.70 \pm 0.007) \,\mu\text{eV}\left(\frac{10^{12}\text{ GeV}}{f_a}\right) \, ,
\end{equation}
where the error includes experimental and theoretical contributions appearing in a detailed calculation of axion-pion mixing~\cite{diCortona:2015ldu}. The constant $f_a$ is related to the scale of spontaneous breaking of the PQ symmetry (see Supplemental Material), and is measured in GeV. What values could this take? The electroweak scale, $\approx 250\text{ GeV}$, was one natural choice, but is excluded experimentally. Other scales in particle physics include the grand unified scale, ${\cal O}(10^{16})\text{ GeV}$, and the Planck scale, $\sim 10^{19}\text{ GeV}$, which is the highest scale where ordinary quantum field theory could possibly remain valid before quantum gravity becomes important. These considerations give only very rough guidance as to the {value of $f_a$ and the} mass of the axion: below we narrow their possible ranges.
\\

The axion mechanism may be implemented in a wide variety of extensions to the Standard Model, which are the subject of much current research~\cite{DiLuzio:2020wdo}. In any axion model, we must introduce a new complex field $\Phi = \chi {\rm e}^{i \theta}$, which gains a non-zero vacuum expectation value that breaks the PQ symmetry spontaneously. After this spontaneous symmetry breaking, the axion is related to the phase of $\Phi$ by $a = N f_a \theta$, where $N$ is an integer, the ``colour anomaly'', which varies depending on the detailed realisation of the Peccei-Quinn mechanism. Figure \ref{fig:potential} is shown for $N=4$. 

For the axion to solve the strong $CP$ problem, the model must also include quarks that are charged under the PQ symmetry, which in turn mediate an interaction between the axion and the gluon force carriers of the strong nuclear force. There are two ways of achieving this. In one class of axion models, the Standard Model quarks are charged under the PQ symmetry \cite{Dine:1981rt,Zhitnitsky:1980tq}: these models have $N=6$. This means we have to add an extra Higgs doublet to the model in order to allow all of the Standard Model interactions to obey the PQ symmetry. In another class of models, we instead add extra heavy, electrically neutral quarks to the theory \cite{PhysRevLett.43.103,SHIFMAN1980493}. Only these extra quarks are charged under the PQ symmetry: the canonical version of this model has $N=1$. The precise values of the axion's mass and interaction strengths depend on these details of the model (see Supplemental Material).

\subsection{Interactions}

The axion does not interact only with quarks and gluons, but also with the other particles in the Standard Model. We know that these interactions must be very weak for the axion to have evaded detection so far.

Particle interactions are governed by their symmetries, and the axion's interactions are set by its {\it pseudoscalar} nature. A pseudoscalar field changes sign under a parity transformation, i.e., when looking at the Universe in a mirror. If we assume that the overall interaction is unchanged by a parity transformation, we find that only certain interactions are allowed for a pseudoscalar particle. These can be expressed (schematically) by the non-relativistic Hamiltonian:

\begin{equation}
    \mathcal{H} = \sqrt{\frac{\epsilon_0}{\mu_0}} g_{a \gamma \gamma} \int a {\bf E} \cdot {\bf B} dV + g_{aff} \hslash c \triangledown a \cdot \hat{{\bf S}} + \sqrt{\epsilon_0 (\hslash c)^3} g_{\rm EDM} a \hat{{\bf S}} \cdot {\bf E},
    \label{eqn:interaction_hamiltonian}
\end{equation}
where $a$ is the axion field measured in units energy; $g_{a \gamma \gamma}$ is the axion's coupling to photons, $\gamma$ (i.e. to electromagnetism), measured in units of inverse energy, $g_{aff}$ is the axion's coupling to matter particles, which are fermions, $f$ (the couplng depends on the particle in question, we write just one case for simplicity); $g_{\rm EDM}$ is the strength of a nuclear EDM induced by the axion; ${\bf E}$ and ${\bf B}$ are the electric and magnetic fields, and $\hat{{\bf S}}$ is the direction of the spin of {\bf the  matter particle in question}. $\epsilon_0$ and $\mu_0$ are the permittivity and permeability of free space respectively - constants associated with electromagnetism, $\hslash$ is the reduced Planck's constant, which parameterizes the size of quantum effects, and $c$ is the speed of light.

These interactions are very different from the interactions of a {\it scalar} particle, which can couple directly ${\bf E}^2 - {\bf B}^2$ {\bf (the scalar Maxwell term)} and to the {\it masses} of matter particles. This means that light scalar fields can mediate extra long-range forces, and their interactions with Standard Model particles are therefore very tightly constrained by the non-observation of such extra forces. In contrast, the couplings of pseudoscalar particles to ${\bf E} \cdot {\bf B}$ and to the {\it spins} of matter particles~\cite{Moody:1984ba,2014PhRvL.113p1801A} make them much harder to detect. \\

What would the axion's interactions with photons and with matter particles look like experimentally? Via its interaction with ${\bf E} \cdot {\bf B}$, axion DM would look like an additional {\it electrical current or anomalous magnetic field}~\cite{1983PhRvL..51.1415S}. More generally, Maxwell's equations of electromagnetism are modified by the addition of the axion field. Further details on how axions modify Maxwell's equations of electromagnetism are discussed in the companion experimental review. Via its coupling to nuclear and electron spins, axion DM would cause these spins to precess, as they would in a magnetic field, but now with an anomalous magnetization caused by the invisible presence of the axion field~\cite{Barbieri:1985cp,2014PhRvX...4b1030B}. 

What are the strengths of the axion's interaction with Standard Model particles? How large are the coupling constants $g_i$ in the Hamiltonian Eq.~\eqref{eqn:interaction_hamiltonian}? We know that the interactions must be very weak, or we would have found axions already. In fact, we expect the axion's couplings to be inversely proportional to the scale of symmetry breaking:
\begin{equation}
    g_i \sim \frac{1}{f_a} \, ,
    \label{eq: couplings}
\end{equation}
This is a consequence of one of the most fundamental ideas in particle physics, namely Effective Field Theory~\cite{Weinberg:1968de}, which tells us to expect the axion's interactions to scale inversely with the energy scale at which the symmetry giving rise to it originates. As this energy $f_a$ could be very high, the axion's interactions could be very weak, as required by the experimental constraints. 
Comparing equations \eqref{eq: mass} and \eqref{eq: couplings}, we see that the axion's couplings obey:
\begin{equation}
    g_i \propto m_a.
    \label{eq: couplings_mass}
\end{equation}
This is true for most models of the QCD axion, whose couplings are generally proportional to its mass, though this relationship can be broken in some specific models of the QCD axion.
Also, as we will see later, the axion is just one particle in the broader class of ALPs discussed in more detail below. Such ALPs need not in general solve the strong $CP$ problem or couple to gluons. This means their mass could take any value and need not be proportional to their couplings, and the constants of proportionality differ wildly for different ALPs. ALPs could therefore be very weakly coupled and hence extremely difficult to detect experimentally. Alternatively, Nature may provide us with more strongly coupled ALPs that can be detected more readily, as discussed below. By contrast, for a canonical QCD axion of a given mass, we can predict the approximate size of the couplings to Standard Model particles, providing a definite target for experimental searches. Further details are given in the supplemental materials. \\

The interaction between DM axions or ALPs with particles and forces in the Standard Model leads to a wide variety of ways to search for them. Initially, in the 1980's up to 2010 or so, experiments were few and far between. There was just one viable method, the microwave cavity \emph{haloscope}, and the axion interactions are too feeble to produce a signal measurable with technology of the time. Now, the landscape has changed. Microwave photon detection and cavity design have allowed the haloscope concept to break ground to exclude regions of the QCD axion parameter space, and expand the search over a wider frequency range~\cite{2010PhRvL.104d1301A,Braine:2019fqb,Semertzidis:2019gkj}. In tandem, new methods to detect axions have been conceived of and developed, including magnetic resonance~\cite{2013PhRvD..88c5023G,2014PhRvX...4b1030B,2018EPJC...78..703C,2019PhRvL.123l1601M,Aybas:2021nvn}, broadband antennas~\cite{2013JCAP...04..016H}, dielectrics and metamaterials~\cite{TheMADMAXWorkingGroup:2016hpc,Egge:2020hyo}, and lumped circuit technology~\cite{Kahn:2016aff,Salemi:2021gck}. These new technologies are at various stages of maturity, with some only existing on paper, others being prototyped, and others already making competitive measurements of axion parameter space and excluding theoretical models. Technologies now in development should be able to cover almost all of the viable axion parameter space in the coming decades: the fuel for the growth of interest in this field. The companion experimental review covers this topic in detail.

As well as searching directly for axion dark matter in the experiments described briefly above and in the companion experimental review, the effects of axions and ALPs may be seen indirectly in telescope observations of stars, galaxies, and galaxy clusters. These astrophysical systems offer extreme environments that would be impossible to replicate on Earth, and are therefore ideal places to search for new physics.

The interaction 
$a {\bf E} \cdot {\bf B}$ means that axions or ALPs can interconvert with photons in the presence of a background magnetic field. This process is mathematically similar to neutrino oscillations, with the crucial difference that the strength of the mixing depends on the size of the external magnetic field. The possibility of interconversion between ALPs and photons in space places the strongest current bounds on $g_{a \gamma \gamma}$ for very low mass ALPs ($m_a \lesssim 10^{-8}$ eV). 

One such bound arises from observations of distant point sources such as Active Galactic Nuclei shining through foreground galaxy clusters. Galaxy clusters host strong magnetic fields over very large distances. If ALPs exist, some of the photons from these point sources would convert into ALPs as they move through the galaxy cluster. From our point of view, some of the light from point sources shining through galaxy clusters would go missing. The non-observation of this effect can be used to place bounds on the axion photon coupling (e.g. \cite{Reynolds:2019uqt}), requiring $g_{a \gamma \gamma} \lesssim 10^{-12} {\,} {\rm GeV}^{-1}$ for ALP masses $m_a \lesssim 10^{-11}$ eV. However, it should be noted that this method relies on the accuracy of our knowledge of the galaxy cluster magnetic field.

We can also bound $g_{a \gamma \gamma}$ using observations of SN 1987A, a supernova observed in 1987 in the nearby (astronomically speaking) Large Magellanic Cloud. A supernova is an explosion at the end of a star's life, which produces vast quantities of neutrinos and photons. If ALPs exist, a supernova would also produce them copiously via nuclear interactions between ALPs and the supernova constituents. Some of the ALPs produced by SN 1987A would have been converted into $\gamma$ - ray photons in the Milky Way's magnetic field. {No such extra $\gamma$ - ray photons have been observed, which} allows us to set a bound $g_{a \gamma \gamma} < 5.3 \times 10^{-12} {\,} {\rm GeV}^{-1}$ for ALP masses $m_a \lesssim 4.4 \times 10^{-10}$ eV (e.g.\cite{Brockway:1996yr}). As we will see below, SN 1987A can also be used to constrain the interactions of higher-mass axions.

The constraints outlined above apply only to rather low mass ALPs. Photons travelling through the plasmas of galaxies and galaxy clusters acquire a low effective mass, but axions or ALPs with masses higher than this effective mass cannot mix efficiently with the photon. For this reason, astrophysical axion-photon mixing constrains the ALP parameter space, but not the standard QCD axion for which $g_{a \gamma \gamma} \propto m_a$.

The coupling between axions and photons would also allow axions to be copiously produced in stars. This would create an additional cooling mechanism for the star and would therefore alter the course of stellar evolution. This effect can also be used to constrain $g_{a \gamma \gamma}$, requiring $g_{a \gamma \gamma} < 6.6 \times 10^{-11} {\,} {\rm GeV}^{-1}$ over a very wide range of masses \cite{Ayala:2014pea}, thus constraining both the QCD axion and a more general ALP. Our nearest star, the Sun, would also copiously produce axions, which could be observed experimentally via controlled axion-photon conversion in the laboratory. This is the aim of the CERN Axion Solar Telescope (CAST) experiment \cite{Anastassopoulos:2017ftl}, and other so-called ``helioscopes'', discussed in the companion experimental review.

\section{Axion Dark Matter Waves}

The advent of quantum mechanics in the early 20th century taught us that for every particle there is a wave, and for every wave there is a particle. The wavelength, $\lambda$, and frequency, $\freq$, of the wave are related to the particle mass $m$ and velocity $v$ (in the non-relativistic limit) by:
\begin{equation}
    \lambda = \frac{h}{m v} \, ,\quad h\freq= E = m c^2+\frac{1}{2}mv^2\, ,
    \label{eqn:wave_DM}
\end{equation}
where $h$ is Planck's constant. When the particle is relatively heavy compared to the axion, as in the case of the electron, the wavelength is small, and so in our day-to-day lives we do not notice the particle behaving as a wave. Electrons are also fermions (particles of half integer spin). This means that they must obey the `exclusion principle' - more than one electron cannot occupy the same state,  and so \emph{collective} wavelike behaviour does not occur (except when they bind to form pairs with integer spin, as in superconductors). 

On the other hand, for light and massless particles, the wavelength of the matter waves can be large, and their frequency low. If these particles are also {\em bosons} (particles of zero or integer spin, see Fig,~\ref{fig:venn}), then many particles can occupy the same state. When the occupation number is macroscopic, the bosons can be described using {\em classical}, rather than quantum, field theory. This gives rise to {\em macroscopic wavelike behaviour}. 
Historically, we were first aware of photons in their guise as the classical electromagnetic radio waves of Hertz. Our technology to observe electromagnetic waves begins in this realm of wavelengths larger than 1 mm or so, and frequencies lower than around 1 THz. 

The axion field, $a$, is just one massive bosonic field among many possibilities, as also is its parent Peccei-Quinn field, $\Phi$. The energy density of dark matter in the Universe is one of the most macroscopic quantities imaginable, which in Einstein's theory of General Relativity affects the very geometry of spacetime and the expansion of the Universe. The theory of axion DM is the theory of classical waves in the fields $\Phi,a$, and their dynamics under the influence of gravity. We begin this section by considering the origin of the energy density in these fields in the very early Universe.\\

\fbox{
\begin{minipage}{5in}
\textbf{What is axion DM and how did it get here?} 
\begin{itemize}
    \item Axion DM consists of the energy stored in spatial and temporal gradients, and potential energy, of the axion field, $a(t,x)$.
    \item Initial fluctuations of the axion field arose from a phase transition involving spontaneous symmetry breaking that occurred in the early Universe.
\end{itemize}
\end{minipage}
}

\subsection{Two Roads to Axion Dark Matter}

Observations of the CMB tell us that to a very good first approximation the early Universe was \emph{homogeneous, isotropic, flat}, and \emph{hot}. Under these approximations, Einstein's equations of General Relativity reduce to a single ordinary differential equation, the Friedmann equation (see Supplementary Material). This equation relates the temperature of the plasma in the early Universe to its expansion rate, expressed in terms of the Hubble parameter $H(t)$. 
\begin{figure}
    \centering
    \includegraphics[width=0.55\textwidth]{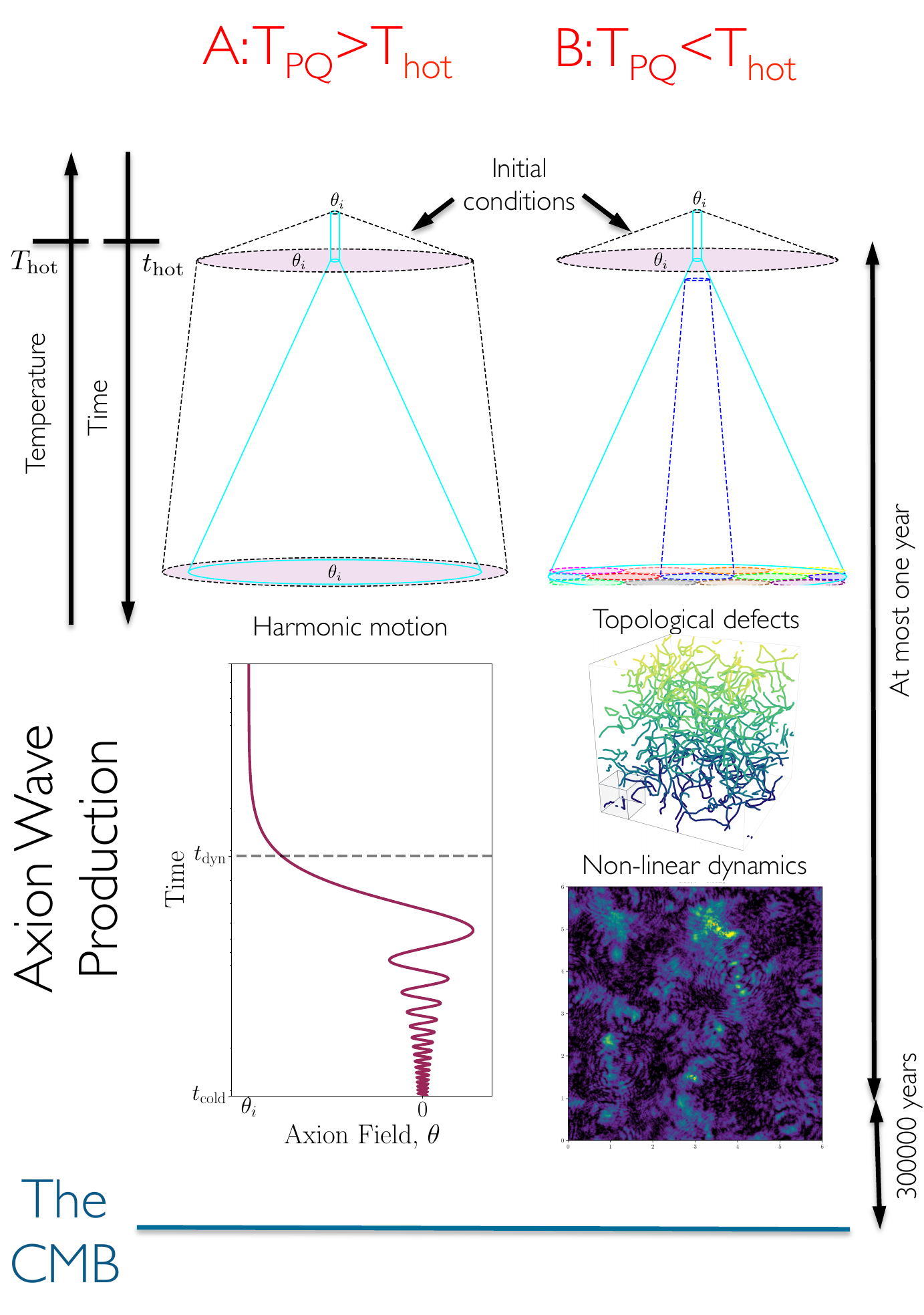}
    \caption{Production of axion DM. {The initial conditions epoch prior to the hot phase serves to smooth the Universe on the largest scales.} The axion dynamics and resulting DM density depends on when the Peccei-Quinn (PQ) phase transition at the temperature $T_\textup{PQ}$ takes place relative to the onset of the hot phase of the evolution of the Universe at $T_\textup{hot}$, leading to two scenarios, A and B.
    Axion production occurs during the period between the two times $t_{\textup{dyn}}$ and $t_{\textup{cold}}$, and is complete in around 1 year at most. After $t_{\textup{cold}}$, axions are described by the Cold Dark Matter model on length scales larger than their de Broglie wavelength. The axions then ``free stream'' up to the time of formation of the Cosmic Microwave Background (CMB), and beyond into the epoch of galaxy formation and the present day. In Scenario A, the axion mass is not fixed by the observed dark matter density, since there is an additional free parameter, $\theta(t_{\rm hot})$. In Scenario B there is no additional free parameter and the mass can in principle be predicted. (Some images adapted from Refs.~\cite{Vaquero:2018tib,2019JCAP...06..047A})}
    \label{fig:two_roads}
\end{figure}

The isotropy of the CMB on the largest scales also provides evidence that this hot, dense state of the early Universe must have been preceded by a far more mysterious epoch that sets the large-scale initial conditions of the Universe. That such an epoch must have existed is the only relevant point at present: which theory describes it (for example inflation or a cyclic Universe) has no bearing on our story (although in the case of inflation, the axion could be used to test key predictions of the theory). After the initial conditions of the Universe were set, the well understood ``hot Big Bang'' phase began, at some finite time $t_{\textup{ hot}}$ when the temperature of the Universe was at its maximum, $T_{\textup{ hot}}$.{When cosmologists refer to the ``age of the Universe'' or ``time after the Big Bang'', this is in reference to the time since $t_{\textup{ hot}}$, and not dependent on how much time there was, or was not, before this (for example, the Universe could be infinitely old in its past). Thus, in the box ``What is known about dark matter?'', the ``Creation of DM'' refers to the amount of time $\Delta t=t-t_{\textup{ hot}}$.} The key distinction for the origin of axion dark matter is whether the Peccei-Quinn phase transition occurs for $T_{\textup{ PQ}}>T_{\textup{ hot}}$ or $T_{\textup{ PQ}}<T_{\textup{ hot}}$. This is summarised in Fig.~\ref{fig:two_roads}. The fact that the oscillations of the axion field caused by spontaneous symmetry breaking leads to a viable DM candidate was first realised in 1983 by three separate groups working independently~\cite{1983PhLB..120..127P,1983PhLB..120..133A,1983PhLB..120..137D}. This happy accident, a DM candidate ``for free'' from a particle postulated only to solve the strong-$CP$ problem, is now a main virtue of the theory. There are two scenarios for spontaneous symmetry breaking that we must consider, shown schematically in Fig.~\ref{fig:two_roads}.

In Scenario A, ($T_{\textup{ PQ}}>T_{\textup{ hot}}$), the Peccei-Quinn phase transition occurs in the epoch during which the initial conditions were set, {\em before} the hot Big Bang. This is the simplest case to describe mathematically, as this epoch serves to smooth the Universe on the largest scales, in accordance with observation. This removes all ``memory'' of the details of the phase transition (e.g., all the spatial gradients in the axion field), and replaces all that complexity with just two quantities: a single random number, the initial value of the axion field, $\theta(t_{\textup{ hot}})$, and the axion mass, $m_a$. With these two inputs, the evolution of the classical axion field, and the subsequent DM density, is determined by a simple wave equation in which the expansion of the Universe acts as a friction term damping the waves (see Supplemental Material).

In Scenario B ($T_{\textup{ PQ}}<T_{\textup{ hot}}$), the Peccei-Quinn phase transition occurred {during the hot Big Bang phase, \emph{after} the initial conditions epoch ended}. {Thus the axion field is not smoothed,} leaving a messy field configuration. The physics involves the Kibble mechanism~\cite{Kibble:1976sj} for the formation and subsequent disposal of ``topological defects'' in the Peccei-Quinn field (the very same mechanism occurs in the superfluid phase transition of Helium~\cite{Zurek:1985qw}). Topological defects are axion field configurations that occur between regions where the axion field takes different values, which may subsequently decay into axion DM. There is also more model dependence in this case, as different models of the axion lead to different kinds of topological defect. The dynamics in Scenario B is highly non-linear. However, in Scenario B the random number, $\theta(t_{\textup{ hot}})$, of Scenario A is averaged over in the initial distribution and is no longer a free parameter. {In Scenario B large density variations in the initial state of the axion field lead to the formation of compact DM objects known as ``miniclusters''~\cite{1988PhLB..205..228H}. These objects offer new opportunities to discover the axion by astrophysical means, but also lead to an increased theoretical uncertainty in the local axion density for DM detection in the laboratory.} 

In both Scenarios A and B an important change happens to the axion field when the expansion rate of the Universe (which is falling during cosmic time) drops below the natural frequency of axion oscillations, which is determined by its mass. At this time, $t_{\textup{dyn}}$, the axion field becomes highly dynamical (the wave equation becomes underdamped) and begins to oscillate. Prior to {$t_\textup{dyn}$}, there were no non-relativistic axions: on the length and time scales of the expansion of the Universe, the axion field was effectively static. This is because the expansion of the universe is responsible for damping the axion field oscillations. 

Thus it is at $t_{\textup{dyn}}$ that we think of axion DM as being produced. In Scenario A, the axion field undergoes damped simple harmonic motion, while in Scenario B, we think of axion field oscillations being emitted from the decaying topological defects. 
This epoch of particle production lasts only for a relatively short period, as the large scale energy density in the axion field, stored up from the phase transition, is converted into field oscillations, which in quantum theory are equivalent to axion particles. From $t_{\textup{dyn}}$ up to some later time $t_{\textup{cold}}$ the number of axion particles grows. Then, from $t_{\textup{cold}}$ up to the present day the number of axion particles in any sufficiently large ``comoving'' volume (a volume that increases in proportion with the expansion of the Universe) in the Universe is conserved , i.e., the axion number density is simply diluted by the expansion of the Universe just like any other particle would be. Axion cold dark matter is born, and starts to form cosmic structures such as galaxies.

\subsection{Cosmic Structure Formation}

Because of the very large number of axions needed to supply the observed DM density,  there are a large number of axions occupying each quantum state. This means that axion DM is described to a very good approximation by the \emph{classical} scalar field $a(t,x)$. Averaged over cosmic scales (distances of more than 10 Megaparsecs), the axion field is very uniform, and its energy density is stored in harmonic oscillations: $\rho=\frac{1}{2}[\dot{a}^2/(\hslash c^3)+m_a^2a^2/(\hslash^3 c^3)]$, with $a=a_0 \cos m_a c^2 t/\hslash$, thus $\rho=m_a^2a_0^2/(2\hslash^3 c^3)$. The value of $a_0$ is determined by the early evolution of the axion field as described above. Crucially, for the axion to behave as cold DM with weak self-interactions, we require $a_0\ll f_a$, which permits the harmonic motion approximation for small displacements. The frequency of oscillation of the field is determined by the particle mass and is given by the Compton frequency.

On smaller scales, gravity causes the axion field to develop inhomogeneities, and it clusters into so-called ``DM halos''. On larger scales, the halos are linked by filaments and sheets in the so-called ``cosmic web''. The process of structure formation begins very early in the history of the Universe: the gravitational potential wells of the DM halos began to form at very high temperatures, when ordinary matter was still completely ionized (i.e., before the CMB formed). The development of inhomogeneities in the axion field is governed by gravity, and the (small) axion self-interactions. The equations describing this evolution are known as the Gross-Pitaevski-Poisson equations, a form of non-linear Schr\"{o}dinger equation for the mean field occupation, where the potential is determined by the density of the field itself via the Poisson equation (see Supplemental Material).

\begin{figure}
    \centering
    \includegraphics[width=0.75\textwidth]{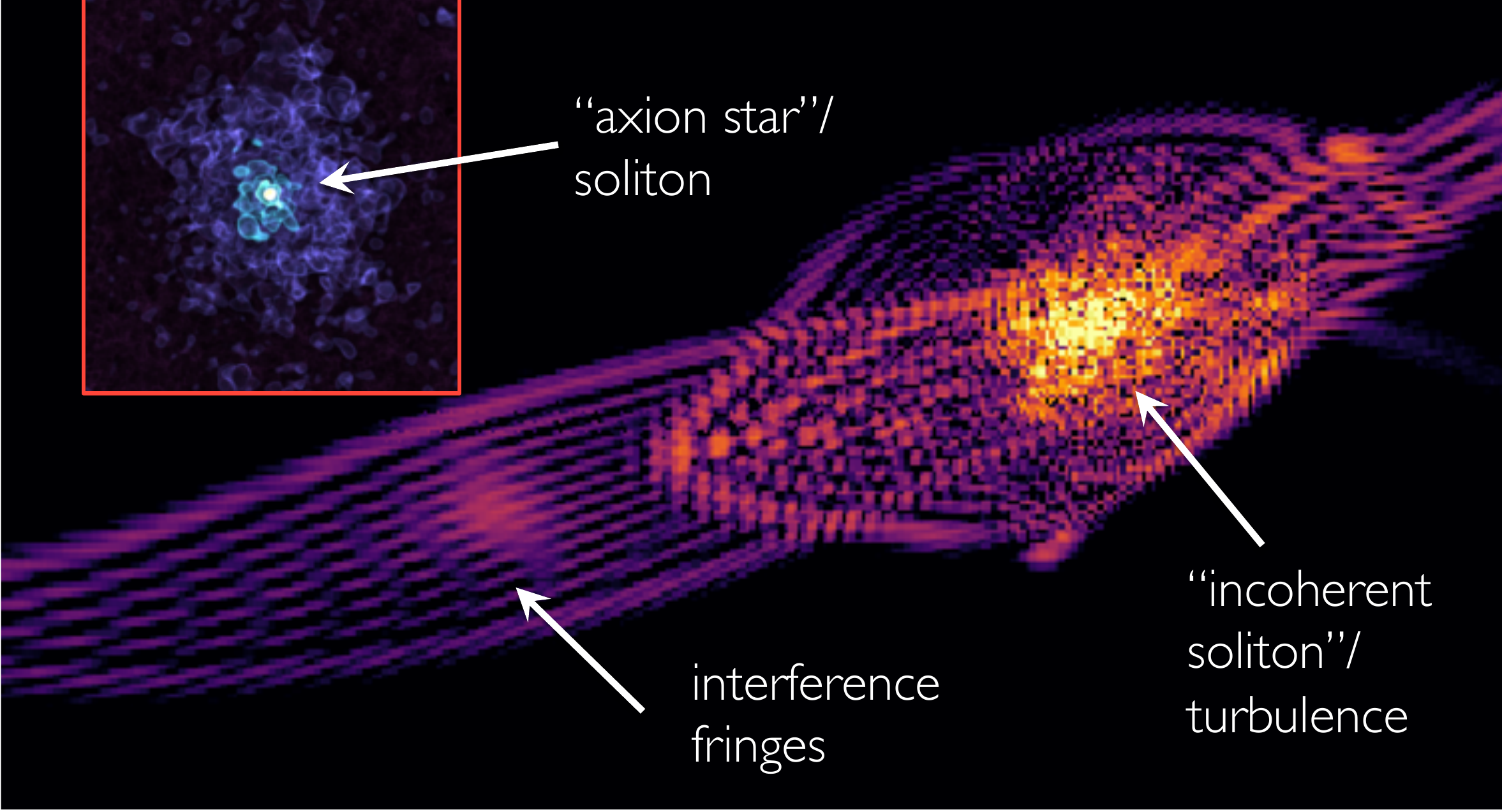}
    \caption{Zoom in to part of the Cosmic Web formed by gravitational interactions and the interference of axion waves. The dense, almost spherical nodes of the web are dark matter halos, which host galaxies like our own. The inset shows a volume rendering of such a halo. Notice the interference of coherent waves outside of dark matter halos forming interference fringes.  Inside halos, there are self bound solitonic objects at high density (``axion stars''), and turbulent waves in the outer regions (``incoherent solitons''). In order to make the unique wave effects visible on cosmic scales (kiloparsecs, kpc), a very low particle mass, $m_a\approx 10^{-22}\text{ eV}$, is used in this simulation. The same physics operates and the same effects occur, however, for all types of bosonic dark matter, with the length scale set by the particle de Broglie wavelength. Figure adapted from work presented in Refs.~\cite{Mocz:2017wlg,Mocz:2019pyf}.}
    \label{fig:fuzzy_dm}
\end{figure}

It is here that axion DM begins to take on some unique characteristics compared to heavier particle DM such as WIMPs. These features are:
\begin{itemize}
    \item The De Broglie wavelength. Gradient energy dominates over gravitational energy on small scales. The axion field is uniform on small scales and there is a minimum mass for a DM halo~\cite{khlopov_scalar}.
    \item Axion Stars. At very high density, the axion forms a kind of soliton (a stable wavepacket-like field configuration) supported by an equilibrium between gravity and gradients~\cite{Seidel:1991zh,Guzman:2006yc}.
    \item Wave turbulence and interference. In the structures of the cosmic web, axion waves have dynamic velocities. Where there are coherent flows, this leads to interference patterns in filaments. In thermalised/virialised environments (halos) the velocities are Maxwell-Boltzmann distributed, and are described by wave turbulence~\cite{2014NatPh..10..496S,Hui:2020hbq}.
\end{itemize}
A numerical simulation of axion structure formation is shown in Fig.~\ref{fig:fuzzy_dm}, where the above mentioned effects can be seen. Similar phenomena showing the formation of solitons and ``incoherent'' turbulent solitons occur in certain non-linear optics systems~\cite{PhysRevLett.107.233901}.

Inside the Milky Way, at the location of the Earth, the axion field is in the turbulent regime. It is given as:
\begin{equation}
    a = \frac{1}{2} \left(\Psi e^{- i m_a c^2 t/\hslash} + \Psi^* e^{+ i m_a c^2 t/\hslash}\right)\,
\end{equation}
The amplitude field $\Psi$ is Rayleigh distributed (up to small corrections) with a coherence length and time governed by the de Broglie wavelength at the local galactic orbital velocity, and an amplitude fixed by the local DM density.
The overall frequency $\freq$ is determined by Eq.~\eqref{eqn:wave_DM}, with the velocity drawn from the Maxwell-Boltzmann distribution, leading to a frequency dispersion $\Delta \freq/\freq\approx 10^{-6}$. The frequency dispersion leads a natural linewidth for axion DM.

\section{What is the Axion Mass?}

\begin{figure}
    \centering
    \includegraphics[width=\textwidth]{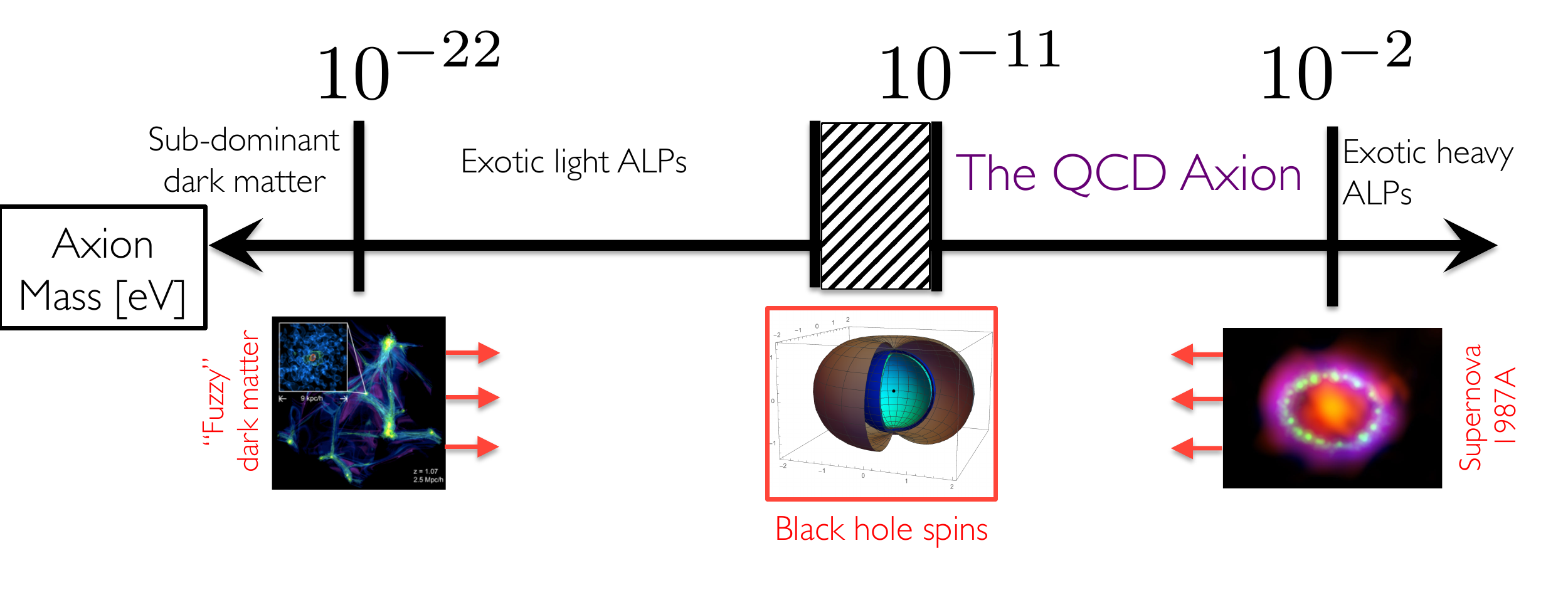}\\
    \caption{Constraints on the axion mass, measured in electronvolts (eV). Red arrows and the shaded region indicate exclusions. The region outside that marked ``The QCD Axion'' corresponds to more exotic axion-like particles (ALPs). ``Fuzzy dark matter'' shows a simulation of cosmic structure formation with very low particle mass DM (adapted with permission from Ref.~\cite{Veltmaat:2018dfz}), at the lower limit of what is acceptable observationally for the dominant DM component. ``Black hole spins'' shows the ergoregion and horizons of spacetime around a spinning black hole, the geometry of which allows very light bosonic particles to extract angular momentum from the black hole. 
    These last two constraints apply under the assumption that axion self-interactions are negligible. The SN1987A constraint applies only to the QCD axion, where the axion-nuclear coupling is proportional to the particle mass, and not to ALPs (image is a composite~\cite{chandra}).}
    \label{fig:axion_mass}
\end{figure}

The QCD axion has a single free parameter in the low-energy theory, the decay constant $f_a$, which in turn sets the axion particle mass (at zero temperature), $m_a$, according to Eq.~\eqref{eq: mass}. In order to try to detect the axion in the laboratory, one would like to know the approximate mass scale, so that experiments can be appropriately targeted to the mean oscillation frequency, $\freq=m_a c^2/ h$, which in turn determines the frequency of any radiation or other associated phenomenon (as discussed elsewhere in this review and in the experimental companion review). Currently, astrophysical considerations provide the best clues to the range of allowed values for the axion mass, summarised in Fig.~\ref{fig:axion_mass}, which we now elaborate on. 

Beginning at the lightest end of the scale are bounds on so-called ``fuzzy'' DM. When the particle mass is very low, the de Broglie wavelength is cosmologically large, and affects the formation and internal dynamics of galaxies. Galaxies cannot form below the de Broglie wavelength, which suppresses galaxy formation in the early Universe (where low-mass galaxies form first). Conservative constraints on the axion mass in this scenario are $m\gtrsim 10^{-22}\text{ eV}$, although the bounds can be tightened by up to three orders of magnitude including more data, and better knowledge of the structure formation process and galactic dynamics~\cite{Marsh:2018zyw,Rogers:2020ltq}. This limit also applies to other forms of bosonic DM (not just axions, see Fig,~\ref{fig:venn}), and is the fundamental lower limit on DM particle mass. If DM is to be lighter than $10^{-22}\text{ eV}$, observationally it can be only a sub-dominant fraction of the total DM density, with current bounds around the 2\% level~\cite{Hlozek:2017zzf}. 

The lack of a significant observed length scale in cosmic structure formation is one of the key facts known about DM, as described in the Box ``What is known about DM?''. Similarly, there is no evidence of a preferred time scale related to the formation of DM, and only a bound on when this must have been complete. Demanding that axions or other ultralight bosons (see next section) were formed early enough sets $t_{\rm cold}\gtrsim 1$ year, and similarly bounds $m\gtrsim 10^{-22}\text{ eV}$.

The next important constraint on the particle mass results from a property known as ``black hole superradiance''. When the Compton wavelength of a boson is resonant with the radius of the ``ergoregion'' of a spinning (Kerr) black hole, the so-called ``Penrose process''~\cite{Penrose:1971uk} operates, and vacuum fluctuations of the boson extract angular momentum from the black hole, reducing its rate of spin. (The ergoregion is a region close to a spinning black hole in which objects are necessarily forced to co-rotate with the black hole.) The radius of the ergoregion is fixed by the black hole mass and spin, while the spin of the black hole can be inferred, for example, from the Doppler shift in the emission of X-rays from its accretion disk. Rapidly spinning black holes would quickly spin down if a resonant boson existed, and so the observation of spinning and old black holes places exclusions on boson masses. The most robust, and also most important, range of these exclusions comes from stellar mass black holes with spins measured from X-ray images, and for spin-0 bosons excludes the range $1\times 10^{-13}\mathrm{\,eV}\leq m\leq 2\times 10^{-11}\mathrm{\,eV}$~\cite{axiverse,Stott:2018opm}, with a similar range excluded for massive spin-1 (vector) bosons. The importance of this bound is that it overlaps the range of the masses allowed for the QCD axion with spontaneous symmetry breaking scale below the Planck scale, $10^{19}\text{ GeV}$. Since it is widely believed that symmetry breaking above the Planck scale is forbidden in quantum gravity and string theory~\cite{Arkani-Hamed:2006emk}, superradiance leads to a solid lower bound to the mass range of the QCD axion.

An upper bound on the QCD axion mass can be derived using the relationship between the mass and the coupling strength to nuclear matter, $g_{aNN}$, and is derived from observations of the supernova SN1987A~\cite{Raffelt:2006cw}. The axion-nucleon coupling leads to emission of axions from neutrons and protons inside a collapsing supernova core in the Bremstrahhlung process, $N+N\rightarrow N+N+a$. The produced axions interact only very weakly with the surrounding nuclear matter, and so can escape from the supernova core, leading to excess cooling compared to the case without axions. Neutrinos are also produced inside supernovae, and provide a known cooling channel. If axions are also produced, then the supernova cools more quickly and thus has less time to emit neutrinos, leading to a lower neutrino flux. The neutrinos produced in SN1987A were observed in neutrino detectors on Earth, and the duration and timing of the neutrino burst is known, and is consistent with predictions of standard supernova models. Thus, the axion production rate must be low, leading to an upper bound on $g_{aNN}$, and thus an upper bound on $m_a$ (see Eq.~\ref{eq: couplings_mass}), which is $m_a\lesssim 2\times 10^{-2}\text{ eV}$.

The superradiance and SN1987A bounds thus limit the QCD axion natural frequency, of critical importance for direct searches, to the range:
\begin{align}
    10\text{ kHz}\lesssim \freq \lesssim 10\text{ THz}\, .
\end{align}
The QCD axion mass can be further narrowed down by appeal to the dark matter density, $\rho_a$, in the two cosmological scenarios discussed in the previous section. Observationally, we have the restriction that the axion density cannot be larger than the observed dark matter density (although it could be smaller, if the axions are not all of the dark matter). 

The important difference between Scenarios A and B is one of predictability. In both cases, the dark matter density depends inversely on the axion mass. In Scenario A, there are two continuous free parameters $\theta_i$ and $m_a$. Thus it is possible to reproduce the correct cosmic density of axions for (almost) any value of the mass $m_a$, with smaller values of the mass having $\theta(t_{\textup{ hot}})\ll 1$. Without a prediction for the axion mass, the experimental task of searching for the axion could be long and arduous. This scenario is probed by experiments operating in the 10 kHz to 10 GHz regime.

In Scenario B, on the other hand, there is only one continuous parameter, $m_a$, and the discrete parameter $N$. In principle, the observed cosmic DM density then predicts the value of $m_a$. The problem with trying to make such a prediction is the computational complexity: the non-linear dynamics of the axion field in Scenario B cannot be solved exactly. Unfortunately, computational limitations even with modern supercomputers prevent a complete, end-to-end, \emph{ab initio} simulation, and the resulting predictions for the mass vary considerably depending on the theoretical methods employed. 

Demanding that the axion is all the DM, the predicted value of the mass is in the range $0.025\text{ meV}\lesssim m_a\lesssim 0.5\text{ meV}$ when $N=1$ (where the range covers approximately the predicted regions from the simulations of Refs.~\cite{Klaer:2017ond,Gorghetto:2020qws}, which span those also of other groups). The case $N>1$ predicts a slightly higher range $0.1\text{ meV}\lesssim m_a\lesssim 20\text{ meV}$ (the upper bound comes from SN1987A)~\cite{Hiramatsu:2012sc}. Lower values of the axion mass are excluded in Scenario B. The theoretical uncertainty is large, but in general such predictions provide motivation to search for axion dark matter using experiments in the 10 GHz to 10 THz frequency range. This promise to \emph{predict} the axion mass uniquely in this scenario is driving many theoretical investigations, and is happening now due to advances in computer simulation techniques.

\section{ALPs Everywhere}

It has been realized since the proposal of the axion that light spin-zero particles with properties resembling those of the axion
may appear in a number of different theoretical contexts. Such axion-like particles (ALPs) may either be pseudoscalar bosons, whose
couplings are odd under parity transformations (reversing spatial directions), or scalar bosons, whose couplings are even under parity. The latter are sometimes
referred to loosely as dilatons or moduli, and may be thought of as lower-mass siblings of the Higgs boson of the Standard Model. 
Generically, both classes of ALPs possess, like axions, couplings with fermion-antifermion pairs,
and also with pairs of photons, gluons, and the other vector bosons
via their field strengths, of the forms ${\bf E}\cdot {\bf B}$ for pseudoscalars and $|{\bf E}|^2+|{\bf B}|^2$ for scalars (where ${\bf E}$ and ${\bf B}$ are generalised for the other forces of the Standard Model). ALPs are distinguished from the QCD axion of our previous discussion, since ALPs do not solve the strong-$CP$ problem. ALPs do, however, still make excellent dark matter candidates.

There is a single scalar Higgs boson in the Standard Model associated with the generation of particle masses via breaking of the 
underlying electroweak symmetry. However, including another symmetry-breaking Higgs field requires the appearance of an additional 
scalar particle as well as a pseudoscalar partner particle. The original axion model is an example of this feature, but it can
be generalized by including more Higgs fields of different types. This scenario occurs in many theories, including supersymmetry, and grand unified theories.

Another theoretical framework where ALPs appear is provided by string theory~\cite{Svrcek:2006yi}. Generic four-dimensional particle physics models
obtained from string theory, by the compactification of extra dimensions of space on small manifolds, contain many additional pseudoscalar 
and scalar fields, which may have (very) small masses. A useful and famous analogy for compact dimensions of space is to imagine our world as a tightrope. From the perspective of large humans, the rope is one-dimensional, and one can only move along its length. However, a small creature such as an ant can circle the ``extra dimension'' of the rope's circumference. In quantum mechanics, length is inversely proportional to energy, and thus the small length scales of extra dimensions of space tend to be associated to large energy scales, possibly even the Planck scale of quantum gravity.  The sizes and shapes of the compact spaces
are unspecified {\em a priori}, since General Relativity and Quantum Mechanics tell us that the compact dimensions must vary in size and shape at different places. These changes in size and shape can be described, from a four-dimensional perspective, by a collection of new {\em fields}, analogous to the Newtonian gravitational field, which in General Relativity describes the curvature of spacetime. Futhermore, there must be some dynamical energy principle that fixes the average size and shape of the compact dimensions; as the potential in Fig.~\ref{fig:potential} fixes the axion angle and also gives it a mass, so these dynamics give masses to the ALPs. This new dynamics is by nature extremely speculative, and (due to the vast array of possibilities for the topology of the compact space) even the most predictive approaches to this problem currently make only statistical statements about the masses of the resulting ALPs. However, the existence of extra dimensions of space is one of the central predictions of string theory and, since it leads to the existence of ALPs, they too can be considered a prediction of the theory.

Although the forms of ALP couplings are similar to those of the original axion, their strengths are quite
model-dependent, and even less is known about them than about the ALP masses. For this reason, there is great interest in designing experimental searches for axions that are sufficiently
general to also be sensitive to wide ranges of the parameter spaces of ALP couplings to both fermions and gauge bosons.

We also mention the growing interest in the possible existence of light vector (spin 1) bosons. In the Standard Model 
there is, on the one hand, the photon whose mass is thought to be exactly zero and is very tightly constrained by experimental
observations, and on the other hand the massive vector bosons responsible for the weak interactions, with no vector
bosons of intermediate mass. However, it is possible that there might be additional light vector bosons that are very
weakly coupled to the particles of the Standard Model, so-called `dark' or `hidden' vector bosons, as also appear
in some string theory models, for example. Their couplings to
Standard Model particles would take different forms from those of ALPs. For example, couplings to pairs of
photons or gluons would be forbidden, and their couplings to fermion-antifermion pairs would be vector- or axial-vector-like.
For this reason, different types of experimental strategies are required to search for them, and we do not discuss them further
in this review.

Finally, we discuss the possibility of axion-like {\em quasiparticles} appearing in certain exotic materials in solid state physics. Quasiparticles such as phonons (quanta of vibrational modes of solids) and magnons (quanta of magnetization) have been staples of condensed matter theory since the earliest days of quantum mechanics. For a quasiparticle to behave as an axion, it must be a pseudoscalar, which requires the parent solid to have certain specific symmetry properties under parity and time-reversal. The would-be axion quasiparicle should also have the trademark axion-like couplings, in particular to ${\bf E}\cdot{\bf B}$, giving rise to axion electrodyanmics. The possibility of axion electrodynamics in solids was identified already by Wilczek, one of the fathers of the axion theory, in the 1980s~\cite{PhysRevLett.58.1799}, but only recently has the quest to realise this idea in the laboratory picked up pace.

Mixing of parallel electric and magnetic fields occurs in general in magnetoelectric materials, and in particular in topological insulators. The existence of the axion quasiparticle requires the breaking of time-reversal symmetry, and is provided by magnetic ions and antiferromagnetic order. Candidate materials to host the axion quasiparticle include Bi(Fe)$_2$Se$_3$ and Mn$_2$Bi$_2$Te$_5$~\cite{2010NatPh...6..284L,Zhang_2020}. The axion quasiparticle has yet to be observed in the lab, although various experiments have been designed to hunt for it, and there is growing interest in these  candidate materials with possible applications to high-speed computer memory~\cite{2018PhLA..382.3018C}. Furthermore, as discussed in the experimental companion to this review, axion quasiparticles could also prove useful in the detection of axion dark matter.

\section{Summary, Conclusions, and Musings}

The dark matter zeitgeist is in flux. For many years experimental and theoretical efforts were directed at WIMPs and similar particle candidates. The technology to detect WIMPs was readily available in the form of the LHC, and the experimental programme of direct WIMP searches also developed rapidly~\cite{Jungman:1995df,Aprile:2018dbl}. Theories attempting to explain the origin of the Higgs mass seemed to predict naturally the existence of WIMP dark matter, with mass {comparable to that} of the Higgs itself, providing a very tantalising prospect. The experimental searches for WIMPs performed excellently, yet to date have found nothing. While there are still important parts of the parameter space left to explore, initial optimism has been {dampened by} experimental cold water.

Almost in parallel to the WIMP story, there was, somewhat behind the scenes, the ever-present and developing story of the axion. Also proposed in the 1970s as a sticking plaster for problems in the Standard Model~\cite{pecceiquinn1977,weinberg1978,wilczek1978}, and with a dark matter candidate for free~\cite{1983PhLB..120..127P,1983PhLB..120..133A,1983PhLB..120..137D}, that is where the shared story of axions and WIMPs ends. The axion is far lighter and more weakly coupled than the WIMP, and initially experimental searches for the axion made slow and painful progress towards the theoretically defined goals. All of this is changing at present and the axion's star is in ascendance. Theorists are working hard trying to hem in the predictions of the axion model, while new technological advances offer prospects for detection in the near future (see the companion experimental review).

Axion dark matter is an ultralight bosonic field, which manifests itself on Earth primarily as exotic oscillating phenomena violating $CP$ symmetry. The primary and defining oscillation is in electric dipole moments, the neutron EDM being of particular interest. The axion also acts as an oscillating source in Maxwell's equations, and can be detected as an anomalous electric or magnetic field. The oscillations in the axion field were set up in the early Universe by a process of symmetry breaking and phase transition, and the discovery of the axion could shed light on those extremely high energy processes in the most distant cosmic past.

The parameter space of axion models is relatively well defined. Astrophysical considerations hem in the oscillation frequency (and thus the particle mass) to a range between kHz and THz, tightening to GHz to THz given some assumptions about the phase transition (see Fig.~\ref{fig:two_roads}). Intense theoretical efforts are being made at present to try and tighten this range still further, limited by computational resources. Experimental searches for the axion often rely on resonant effects~\cite{1983PhRvL..51.1415S}, and it is the task of tuning these experimental ``radios'' over the wide range of available frequencies that makes axion searches so long and laborious. If the theorists could predict the axion mass accurately, which might be within reach, the experimental search could be tuned, and accomplished within a very short period. However, we should temper our optimism. The theoretical predictions rely on a large number of assumptions, which could be violated in Nature. 

Axion theorists are also intensely engaged in thinking about the axion couplings, $g$, to ordinary matter~\cite{DiLuzio:2020wdo}. Astrophysical constraints, for example from stellar evolution, often provide a good guide for how large these couplings can be in general, without theoretical bias. The axion theory predicts $g\propto m_a$, but the constant of proportionality varies for different models. A given experiment or observation constrains the value of the constant of proportionality, and it is for theorists to interpret which, if any, of the available axion models has been excluded. Again, efforts to pin down the range of allowed couplings are starting to reach some degree of consensus, just in time for the data onslaught that is coming our way from new experiments.

As described in the companion experimental review, this onslaught of data is already beginning, and should reach a conclusion in the next decade or so. Experiments with resonant circuits and nuclear magnetic resonance operate in the kHz range and below, microwave cavities in the GHz range, and the use of dielectrics and novel materials can push into near THz and above. If these experiments succeed, the next task will be to identify the axion precisely. One would like to measure the axion via all its interactions, Eq.~\eqref{eqn:interaction_hamiltonian}. Fortunately, once one experiment succeeds and finds the right axion ``radio station'' (i.e. the value of the axion mass), all the others will quickly be able to tune in and measure the different couplings of the axion. For example, with the axion photon and electron couplings in hand, it should be possible to determine whether the axion speaks to the Standard Model via new heavy quarks~\cite{PhysRevLett.43.103,SHIFMAN1980493}, or via the Higgs boson~\cite{Dine:1981rt,Zhitnitsky:1980tq} (see Supplemental Material).

An axion dark matter experiment in reality measures a combination of the axion coupling and the dark matter density, so another key task will be to use overlapping experiments that do not rely on the dark matter density to break this degeneracy. Fortunately, ideas are available using nuclear magnetic resonance, electric dipole measurements, searches for new forces, and combinations of searches for axions from the Sun and other astrophysical sources (see the experimental companion to this review). Axion searches also offer a wealth of information about the frequency structure of the dark matter, which would provide a detailed view of dark matter in our galactic neighbourhood, and open a new field of dark matter astronomy~\cite{OHare:2017yze}. This combination of astrophysics and particle physics could eventually determine not just if the axion is indeed the dark matter, but exactly what fraction of the dark matter it is and how it is distributed. If the axion is just one part of the dark matter, a new series of more challenging problems arise, since all dark matter searches are harder when aimed at sub-components.

Fortunately, theorists are prepared for such an eventuality. Many theories possessing the axion also possess a zoo of related ALPs, all or none of which could also contribute to the dark matter, and some of which could even be related to hidden dimensions of spacetime~\cite{Svrcek:2006yi}. If the axion is discovered, it will open up the doors to experimentally testing these far reaching theories. The interest in these related particles, and experimental searches for them, is also reaching new heights, but that is a story too varied for this review.

The story of the axion, like much of modern particle physics, began in the 1970s, rooted in ideas of symmetry. The Standard Model is a symmetry success story, but other theories based on symmetry have failed. Does Nature really play by these rules? The popularity of the axion has risen as a prime candidate to explain dark matter, the cosmic glue that in many ways is responsible for our own existence. What composes this mysterious substance, and how did it get here? The discovery of the axion seems tantalisingly close, promising answers to these deepest of questions.

\section*{Acknowledgements}

{\it We were saddened to hear during the completion of this paper of the deaths of Toshihide Maskawa and Steven Weinberg, whose deep theoretical insights underlie the topics reviewed here.}\\
~~\\
We are indebted to Marco Gorghetto for providing an image used in Fig.~\ref{fig:two_roads}, to Matthew Stott for providing an image used in Fig.~\ref{fig:axion_mass}, and to Philip Mocz who provided the images used in Fig.~\ref{fig:fuzzy_dm}, and to all authors who have allowed us to reproduce images from their published work. The Chandra composite image in Fig.~\ref{fig:axion_mass} has credit to the following sources, X-ray: NASA/CXC/SAO/PSU \cite{Frank:2016qwj}; Optical: NASA/STScI \cite{Orlando:2015dfa}; Millimeter: ESO/NAOJ/NRAO/ALMA \cite{Indebetouw:2013sva}. We thank Yannis Smertzidis (CAPP), Ian Williams (University of Surrey), and SungWoo Youn (CAPP) for comments on the draft. 
FC-D is supported by the UK STFC grant ST/P001246/1 and Stephen  Hawking Fellowship EP/T01668X/1.
The work of JE was supported in part by the UK STFC via grant ST/T000759/1, and in part by the Estonian Research Council via grant MOBTT5. DJEM was previously supported by the Alexander von Humboldt Foundation and the German Federal Ministry of Education and Research, and is now supported by a UK STFC Ernest Rutherford Fellowship. 
\subsection*{Data and Materials Availability}
All data needed to evaluate the conclusions in the paper are present in the paper and/or the Supplementary Materials.
\subsection*{Author Contributions}
All authors contributed equally to the writing and editing of the manuscript.
\subsection*{Competing Interests}
The authors declare no competing interests.

\bibliographystyle{naturemag}
\bibliography{new_bib}

\begin{thebibliography}{10}
\expandafter\ifx\csname url\endcsname\relax
  \def\url#1{\texttt{#1}}\fi
\expandafter\ifx\csname urlprefix\endcsname\relax\def\urlprefix{URL }\fi
\providecommand{\bibinfo}[2]{#2}
\providecommand{\eprint}[2][]{\url{#2}}

\bibitem{ATLASH}
\bibinfo{author}{Aad, G.} \emph{et~al.}
\newblock \bibinfo{title}{{Observation of a new particle in the search for the
  Standard Model Higgs boson with the ATLAS detector at the LHC}}.
\newblock \emph{\bibinfo{journal}{Phys. Lett. B}}
  \textbf{\bibinfo{volume}{716}}, \bibinfo{pages}{1--29}
  (\bibinfo{year}{2012}).
\newblock \eprint{1207.7214}.

\bibitem{CMSH}
\bibinfo{author}{Chatrchyan, S.} \emph{et~al.}
\newblock \bibinfo{title}{{Observation of a New Boson at a Mass of 125 GeV with
  the CMS Experiment at the LHC}}.
\newblock \emph{\bibinfo{journal}{Phys. Lett. B}}
  \textbf{\bibinfo{volume}{716}}, \bibinfo{pages}{30--61}
  (\bibinfo{year}{2012}).
\newblock \eprint{1207.7235}.

\bibitem{Higgs:1964pj}
\bibinfo{author}{Higgs, P.~W.}
\newblock \bibinfo{title}{{Broken Symmetries and the Masses of Gauge Bosons}}.
\newblock \emph{\bibinfo{journal}{Phys. Rev. Lett.}}
  \textbf{\bibinfo{volume}{13}}, \bibinfo{pages}{508--509}
  (\bibinfo{year}{1964}).

\bibitem{Englert:1964et}
\bibinfo{author}{Englert, F.} \& \bibinfo{author}{Brout, R.}
\newblock \bibinfo{title}{{Broken Symmetry and the Mass of Gauge Vector
  Mesons}}.
\newblock \emph{\bibinfo{journal}{Phys. Rev. Lett.}}
  \textbf{\bibinfo{volume}{13}}, \bibinfo{pages}{321--323}
  (\bibinfo{year}{1964}).

\bibitem{pecceiquinn1977}
\bibinfo{author}{{Peccei}, R.~D.} \& \bibinfo{author}{{Quinn}, H.~R.}
\newblock \bibinfo{title}{{CP conservation in the presence of
  pseudoparticles}}.
\newblock \emph{\bibinfo{journal}{Phys. Rev. Lett.}}
  \textbf{\bibinfo{volume}{38}}, \bibinfo{pages}{1440--1443}
  (\bibinfo{year}{1977}).

\bibitem{weinberg1978}
\bibinfo{author}{{Weinberg}, S.}
\newblock \bibinfo{title}{{A new light boson?}}
\newblock \emph{\bibinfo{journal}{Phys. Rev. Lett.}}
  \textbf{\bibinfo{volume}{40}}, \bibinfo{pages}{223--226}
  (\bibinfo{year}{1978}).

\bibitem{wilczek1978}
\bibinfo{author}{{Wilczek}, F.}
\newblock \bibinfo{title}{{Problem of strong P and T invariance in the presence
  of instantons}}.
\newblock \emph{\bibinfo{journal}{Phys. Rev. Lett.}}
  \textbf{\bibinfo{volume}{40}}, \bibinfo{pages}{279--282}
  (\bibinfo{year}{1978}).

\bibitem{Dirac:1928hu}
\bibinfo{author}{Dirac, P. A.~M.}
\newblock \bibinfo{title}{{The quantum theory of the electron}}.
\newblock \emph{\bibinfo{journal}{Proc. Roy. Soc. Lond. A}}
  \textbf{\bibinfo{volume}{117}}, \bibinfo{pages}{610--624}
  (\bibinfo{year}{1928}).

\bibitem{Sakharov:1967dj}
\bibinfo{author}{Sakharov, A.~D.}
\newblock \bibinfo{title}{{Violation of CP Invariance, C asymmetry, and baryon
  asymmetry of the universe}}.
\newblock \emph{\bibinfo{journal}{Pisma Zh. Eksp. Teor. Fiz.}}
  \textbf{\bibinfo{volume}{5}}, \bibinfo{pages}{32--35} (\bibinfo{year}{1967}).

\bibitem{Christenson:1964fg}
\bibinfo{author}{Christenson, J.~H.}, \bibinfo{author}{Cronin, J.~W.},
  \bibinfo{author}{Fitch, V.~L.} \& \bibinfo{author}{Turlay, R.}
\newblock \bibinfo{title}{{Evidence for the $2\pi$ Decay of the $K_2^0$
  Meson}}.
\newblock \emph{\bibinfo{journal}{Phys. Rev. Lett.}}
  \textbf{\bibinfo{volume}{13}}, \bibinfo{pages}{138--140}
  (\bibinfo{year}{1964}).

\bibitem{Kobayashi:1973fv}
\bibinfo{author}{Kobayashi, M.} \& \bibinfo{author}{Maskawa, T.}
\newblock \bibinfo{title}{{CP Violation in the Renormalizable Theory of Weak
  Interaction}}.
\newblock \emph{\bibinfo{journal}{Prog. Theor. Phys.}}
  \textbf{\bibinfo{volume}{49}}, \bibinfo{pages}{652--657}
  (\bibinfo{year}{1973}).

\bibitem{Rubin:1970zza}
\bibinfo{author}{Rubin, V.~C.} \& \bibinfo{author}{Ford, W.~K., Jr.}
\newblock \bibinfo{title}{{Rotation of the Andromeda Nebula from a
  Spectroscopic Survey of Emission Regions}}.
\newblock \emph{\bibinfo{journal}{Astrophys. J.}}
  \textbf{\bibinfo{volume}{159}}, \bibinfo{pages}{379--403}
  (\bibinfo{year}{1970}).

\bibitem{Aghanim:2018eyx}
\bibinfo{author}{Aghanim, N.} \emph{et~al.}
\newblock \bibinfo{title}{{Planck 2018 results. VI. Cosmological parameters}}.
\newblock \emph{\bibinfo{journal}{Astron. Astrophys.}}
  \textbf{\bibinfo{volume}{641}}, \bibinfo{pages}{A6} (\bibinfo{year}{2020}).
\newblock \eprint{1807.06209}.

\bibitem{Peebles:1982ff}
\bibinfo{author}{Peebles, P. J.~E.}
\newblock \bibinfo{title}{{Large scale background temperature and mass
  fluctuations due to scale invariant primeval perturbations}}.
\newblock \emph{\bibinfo{journal}{Astrophys. J. Lett.}}
  \textbf{\bibinfo{volume}{263}}, \bibinfo{pages}{L1--L5}
  (\bibinfo{year}{1982}).

\bibitem{Jungman:1995df}
\bibinfo{author}{Jungman, G.}, \bibinfo{author}{Kamionkowski, M.} \&
  \bibinfo{author}{Griest, K.}
\newblock \bibinfo{title}{{Supersymmetric dark matter}}.
\newblock \emph{\bibinfo{journal}{Phys. Rept.}} \textbf{\bibinfo{volume}{267}},
  \bibinfo{pages}{195--373} (\bibinfo{year}{1996}).
\newblock \eprint{hep-ph/9506380}.

\bibitem{Aprile:2018dbl}
\bibinfo{author}{Aprile, E.} \emph{et~al.}
\newblock \bibinfo{title}{{Dark Matter Search Results from a One Ton-Year
  Exposure of XENON1T}}.
\newblock \emph{\bibinfo{journal}{Phys. Rev. Lett.}}
  \textbf{\bibinfo{volume}{121}}, \bibinfo{pages}{111302}
  (\bibinfo{year}{2018}).
\newblock \eprint{1805.12562}.

\bibitem{Clowe:2003tk}
\bibinfo{author}{Clowe, D.}, \bibinfo{author}{Gonzalez, A.} \&
  \bibinfo{author}{Markevitch, M.}
\newblock \bibinfo{title}{{Weak lensing mass reconstruction of the interacting
  cluster 1E0657-558: Direct evidence for the existence of dark matter}}.
\newblock \emph{\bibinfo{journal}{Astrophys. J.}}
  \textbf{\bibinfo{volume}{604}}, \bibinfo{pages}{596--603}
  (\bibinfo{year}{2004}).
\newblock \eprint{astro-ph/0312273}.

\bibitem{Abel:2020gbr}
\bibinfo{author}{Abel, C.} \emph{et~al.}
\newblock \bibinfo{title}{{Measurement of the permanent electric dipole moment
  of the neutron}}.
\newblock \emph{\bibinfo{journal}{Phys. Rev. Lett.}}
  \textbf{\bibinfo{volume}{124}}, \bibinfo{pages}{081803}
  (\bibinfo{year}{2020}).
\newblock \eprint{2001.11966}.

\bibitem{Nambu:1960xd}
\bibinfo{author}{Nambu, Y.}
\newblock \bibinfo{title}{{Axial vector current conservation in weak
  interactions}}.
\newblock \emph{\bibinfo{journal}{Phys. Rev. Lett.}}
  \textbf{\bibinfo{volume}{4}}, \bibinfo{pages}{380--382}
  (\bibinfo{year}{1960}).

\bibitem{Goldstone:1961eq}
\bibinfo{author}{Goldstone, J.}
\newblock \bibinfo{title}{{Field Theories with Superconductor Solutions}}.
\newblock \emph{\bibinfo{journal}{Nuovo Cim.}} \textbf{\bibinfo{volume}{19}},
  \bibinfo{pages}{154--164} (\bibinfo{year}{1961}).

\bibitem{diCortona:2015ldu}
\bibinfo{author}{Grilli~di Cortona, G.}, \bibinfo{author}{Hardy, E.},
  \bibinfo{author}{Pardo~Vega, J.} \& \bibinfo{author}{Villadoro, G.}
\newblock \bibinfo{title}{{The QCD axion, precisely}}.
\newblock \emph{\bibinfo{journal}{JHEP}} \textbf{\bibinfo{volume}{01}},
  \bibinfo{pages}{034} (\bibinfo{year}{2016}).
\newblock \eprint{1511.02867}.

\bibitem{DiLuzio:2020wdo}
\bibinfo{author}{Di~Luzio, L.}, \bibinfo{author}{Giannotti, M.},
  \bibinfo{author}{Nardi, E.} \& \bibinfo{author}{Visinelli, L.}
\newblock \bibinfo{title}{{The landscape of QCD axion models}}.
\newblock \emph{\bibinfo{journal}{Phys. Rept.}} \textbf{\bibinfo{volume}{870}},
  \bibinfo{pages}{1--117} (\bibinfo{year}{2020}).
\newblock \eprint{2003.01100}.

\bibitem{Dine:1981rt}
\bibinfo{author}{Dine, M.}, \bibinfo{author}{Fischler, W.} \&
  \bibinfo{author}{Srednicki, M.}
\newblock \bibinfo{title}{{A Simple Solution to the Strong CP Problem with a
  Harmless Axion}}.
\newblock \emph{\bibinfo{journal}{Phys. Lett. B}}
  \textbf{\bibinfo{volume}{104}}, \bibinfo{pages}{199--202}
  (\bibinfo{year}{1981}).

\bibitem{Zhitnitsky:1980tq}
\bibinfo{author}{Zhitnitsky, A.~R.}
\newblock \bibinfo{title}{{On Possible Suppression of the Axion Hadron
  Interactions. (In Russian)}}.
\newblock \emph{\bibinfo{journal}{Sov. J. Nucl. Phys.}}
  \textbf{\bibinfo{volume}{31}}, \bibinfo{pages}{260} (\bibinfo{year}{1980}).

\bibitem{PhysRevLett.43.103}
\bibinfo{author}{Kim, J.~E.}
\newblock \bibinfo{title}{Weak-interaction singlet and strong $\mathrm{CP}$
  invariance}.
\newblock \emph{\bibinfo{journal}{Phys. Rev. Lett.}}
  \textbf{\bibinfo{volume}{43}}, \bibinfo{pages}{103--107}
  (\bibinfo{year}{1979}).
\newblock \urlprefix\url{https://link.aps.org/doi/10.1103/PhysRevLett.43.103}.

\bibitem{SHIFMAN1980493}
\bibinfo{author}{Shifman, M.}, \bibinfo{author}{Vainshtein, A.} \&
  \bibinfo{author}{Zakharov, V.}
\newblock \bibinfo{title}{Can confinement ensure natural cp invariance of
  strong interactions?}
\newblock \emph{\bibinfo{journal}{Nuclear Physics B}}
  \textbf{\bibinfo{volume}{166}}, \bibinfo{pages}{493--506}
  (\bibinfo{year}{1980}).
\newblock
  \urlprefix\url{https://www.sciencedirect.com/science/article/pii/0550321380902096}.

\bibitem{Moody:1984ba}
\bibinfo{author}{Moody, J.} \& \bibinfo{author}{Wilczek, F.}
\newblock \bibinfo{title}{{NEW MACROSCOPIC FORCES?}}
\newblock \emph{\bibinfo{journal}{\prd}} \textbf{\bibinfo{volume}{30}},
  \bibinfo{pages}{130} (\bibinfo{year}{1984}).

\bibitem{2014PhRvL.113p1801A}
\bibinfo{author}{{Arvanitaki}, A.} \& \bibinfo{author}{{Geraci}, A.~A.}
\newblock \bibinfo{title}{{Resonantly Detecting Axion-Mediated Forces with
  Nuclear Magnetic Resonance}}.
\newblock \emph{\bibinfo{journal}{Phys. Rev. Lett.}}
  \textbf{\bibinfo{volume}{113}}, \bibinfo{pages}{161801}
  (\bibinfo{year}{2014}).
\newblock \eprint{1403.1290}.

\bibitem{1983PhRvL..51.1415S}
\bibinfo{author}{{Sikivie}, P.}
\newblock \bibinfo{title}{{Experimental tests of the 'invisible' axion}}.
\newblock \emph{\bibinfo{journal}{Phys. Rev. Lett.}}
  \textbf{\bibinfo{volume}{51}}, \bibinfo{pages}{1415--1417}
  (\bibinfo{year}{1983}).

\bibitem{Barbieri:1985cp}
\bibinfo{author}{Barbieri, R.}, \bibinfo{author}{Cerdonio, M.},
  \bibinfo{author}{Fiorentini, G.} \& \bibinfo{author}{Vitale, S.}
\newblock \bibinfo{title}{{AXION TO MAGNON CONVERSION: A SCHEME FOR THE
  DETECTION OF GALACTIC AXIONS}}.
\newblock \emph{\bibinfo{journal}{Phys. Lett. B}}
  \textbf{\bibinfo{volume}{226}}, \bibinfo{pages}{357--360}
  (\bibinfo{year}{1989}).

\bibitem{2014PhRvX...4b1030B}
\bibinfo{author}{{Budker}, D.}, \bibinfo{author}{{Graham}, P.~W.},
  \bibinfo{author}{{Ledbetter}, M.}, \bibinfo{author}{{Rajendran}, S.} \&
  \bibinfo{author}{{Sushkov}, A.~O.}
\newblock \bibinfo{title}{{Proposal for a Cosmic Axion Spin Precession
  Experiment (CASPEr)}}.
\newblock \emph{\bibinfo{journal}{Phys. Rev. X}} \textbf{\bibinfo{volume}{4}},
  \bibinfo{pages}{021030} (\bibinfo{year}{2014}).
\newblock \eprint{1306.6089}.

\bibitem{Weinberg:1968de}
\bibinfo{author}{Weinberg, S.}
\newblock \bibinfo{title}{{Nonlinear realizations of chiral symmetry}}.
\newblock \emph{\bibinfo{journal}{Phys. Rev.}} \textbf{\bibinfo{volume}{166}},
  \bibinfo{pages}{1568--1577} (\bibinfo{year}{1968}).

\bibitem{2010PhRvL.104d1301A}
\bibinfo{author}{{Asztalos}, S.~J.} \emph{et~al.}
\newblock \bibinfo{title}{{SQUID-Based Microwave Cavity Search for Dark-Matter
  Axions}}.
\newblock \emph{\bibinfo{journal}{Phys. Rev. Lett.}}
  \textbf{\bibinfo{volume}{104}}, \bibinfo{pages}{041301}
  (\bibinfo{year}{2010}).
\newblock \eprint{0910.5914}.

\bibitem{Braine:2019fqb}
\bibinfo{author}{Braine, T.} \emph{et~al.}
\newblock \bibinfo{title}{{Extended Search for the Invisible Axion with the
  Axion Dark Matter Experiment}}.
\newblock \emph{\bibinfo{journal}{Phys. Rev. Lett.}}
  \textbf{\bibinfo{volume}{124}}, \bibinfo{pages}{101303}
  (\bibinfo{year}{2020}).
\newblock \eprint{1910.08638}.

\bibitem{Semertzidis:2019gkj}
\bibinfo{author}{Semertzidis, Y.~K.} \emph{et~al.}
\newblock \bibinfo{title}{{Axion Dark Matter Research with IBS/CAPP}}
  (\bibinfo{year}{2019}).
\newblock \eprint{1910.11591}.

\bibitem{2013PhRvD..88c5023G}
\bibinfo{author}{{Graham}, P.~W.} \& \bibinfo{author}{{Rajendran}, S.}
\newblock \bibinfo{title}{{New observables for direct detection of axion dark
  matter}}.
\newblock \emph{\bibinfo{journal}{\prd}} \textbf{\bibinfo{volume}{88}},
  \bibinfo{pages}{035023} (\bibinfo{year}{2013}).
\newblock \eprint{1306.6088}.

\bibitem{2018EPJC...78..703C}
\bibinfo{author}{{Crescini}, N.} \emph{et~al.}
\newblock \bibinfo{title}{{Operation of a ferromagnetic axion haloscope at
  m\_a=58 {\ensuremath{\mu}} eV}}.
\newblock \emph{\bibinfo{journal}{European Physical Journal C}}
  \textbf{\bibinfo{volume}{78}}, \bibinfo{pages}{703} (\bibinfo{year}{2018}).
\newblock \eprint{1806.00310}.

\bibitem{2019PhRvL.123l1601M}
\bibinfo{author}{{Marsh}, D. J.~E.}, \bibinfo{author}{{Fong}, K.~C.},
  \bibinfo{author}{{Lentz}, E.~W.}, \bibinfo{author}{{{\v{S}}mejkal}, L.} \&
  \bibinfo{author}{{Ali}, M.~N.}
\newblock \bibinfo{title}{{Proposal to Detect Dark Matter using Axionic
  Topological Antiferromagnets}}.
\newblock \emph{\bibinfo{journal}{Phys. Rev. Lett.}}
  \textbf{\bibinfo{volume}{123}}, \bibinfo{pages}{121601}
  (\bibinfo{year}{2019}).
\newblock \eprint{1807.08810}.

\bibitem{Aybas:2021nvn}
\bibinfo{author}{Aybas, D.} \emph{et~al.}
\newblock \bibinfo{title}{{Search for Axionlike Dark Matter Using Solid-State
  Nuclear Magnetic Resonance}}.
\newblock \emph{\bibinfo{journal}{Phys. Rev. Lett.}}
  \textbf{\bibinfo{volume}{126}}, \bibinfo{pages}{141802}
  (\bibinfo{year}{2021}).
\newblock \eprint{2101.01241}.

\bibitem{2013JCAP...04..016H}
\bibinfo{author}{{Horns}, D.} \emph{et~al.}
\newblock \bibinfo{title}{{Searching for WISPy cold dark matter with a dish
  antenna}}.
\newblock \emph{\bibinfo{journal}{\jcap}} \textbf{\bibinfo{volume}{2013}},
  \bibinfo{pages}{016} (\bibinfo{year}{2013}).
\newblock \eprint{1212.2970}.

\bibitem{TheMADMAXWorkingGroup:2016hpc}
\bibinfo{author}{Caldwell, A.} \emph{et~al.}
\newblock \bibinfo{title}{{Dielectric Haloscopes: A New Way to Detect Axion
  Dark Matter}}.
\newblock \emph{\bibinfo{journal}{Phys. Rev. Lett.}}
  \textbf{\bibinfo{volume}{118}}, \bibinfo{pages}{091801}
  (\bibinfo{year}{2017}).
\newblock \eprint{1611.05865}.

\bibitem{Egge:2020hyo}
\bibinfo{author}{Egge, J.}, \bibinfo{author}{Knirck, S.},
  \bibinfo{author}{Majorovits, B.}, \bibinfo{author}{Moore, C.} \&
  \bibinfo{author}{Reimann, O.}
\newblock \bibinfo{title}{{A first proof of principle booster setup for the
  MADMAX dielectric haloscope}}.
\newblock \emph{\bibinfo{journal}{Eur. Phys. J. C}}
  \textbf{\bibinfo{volume}{80}}, \bibinfo{pages}{392} (\bibinfo{year}{2020}).
\newblock \eprint{2001.04363}.

\bibitem{Kahn:2016aff}
\bibinfo{author}{Kahn, Y.}, \bibinfo{author}{Safdi, B.~R.} \&
  \bibinfo{author}{Thaler, J.}
\newblock \bibinfo{title}{{Broadband and Resonant Approaches to Axion Dark
  Matter Detection}}.
\newblock \emph{\bibinfo{journal}{Phys. Rev. Lett.}}
  \textbf{\bibinfo{volume}{117}}, \bibinfo{pages}{141801}
  (\bibinfo{year}{2016}).
\newblock \eprint{1602.01086}.

\bibitem{Salemi:2021gck}
\bibinfo{author}{Salemi, C.~P.} \emph{et~al.}
\newblock \bibinfo{title}{{Search for Low-Mass Axion Dark Matter with
  ABRACADABRA-10~cm}}.
\newblock \emph{\bibinfo{journal}{Phys. Rev. Lett.}}
  \textbf{\bibinfo{volume}{127}}, \bibinfo{pages}{081801}
  (\bibinfo{year}{2021}).
\newblock \eprint{2102.06722}.

\bibitem{Reynolds:2019uqt}
\bibinfo{author}{Reynolds, C.~S.} \emph{et~al.}
\newblock \bibinfo{title}{{Astrophysical limits on very light axion-like
  particles from Chandra grating spectroscopy of NGC 1275}}
  (\bibinfo{year}{2019}).
\newblock \eprint{1907.05475}.

\bibitem{Brockway:1996yr}
\bibinfo{author}{Brockway, J.~W.}, \bibinfo{author}{Carlson, E.~D.} \&
  \bibinfo{author}{Raffelt, G.~G.}
\newblock \bibinfo{title}{{SN1987A gamma-ray limits on the conversion of
  pseudoscalars}}.
\newblock \emph{\bibinfo{journal}{Phys. Lett. B}}
  \textbf{\bibinfo{volume}{383}}, \bibinfo{pages}{439--443}
  (\bibinfo{year}{1996}).
\newblock \eprint{astro-ph/9605197}.

\bibitem{Ayala:2014pea}
\bibinfo{author}{Ayala, A.}, \bibinfo{author}{Dom\'\i{}nguez, I.},
  \bibinfo{author}{Giannotti, M.}, \bibinfo{author}{Mirizzi, A.} \&
  \bibinfo{author}{Straniero, O.}
\newblock \bibinfo{title}{{Revisiting the bound on axion-photon coupling from
  Globular Clusters}}.
\newblock \emph{\bibinfo{journal}{Phys. Rev. Lett.}}
  \textbf{\bibinfo{volume}{113}}, \bibinfo{pages}{191302}
  (\bibinfo{year}{2014}).
\newblock \eprint{1406.6053}.

\bibitem{Anastassopoulos:2017ftl}
\bibinfo{author}{Anastassopoulos, V.} \emph{et~al.}
\newblock \bibinfo{title}{{New CAST Limit on the Axion-Photon Interaction}}.
\newblock \emph{\bibinfo{journal}{Nature Phys.}} \textbf{\bibinfo{volume}{13}},
  \bibinfo{pages}{584--590} (\bibinfo{year}{2017}).
\newblock \eprint{1705.02290}.

\bibitem{Vaquero:2018tib}
\bibinfo{author}{Vaquero, A.}, \bibinfo{author}{Redondo, J.} \&
  \bibinfo{author}{Stadler, J.}
\newblock \bibinfo{title}{{Early seeds of axion miniclusters}}.
\newblock \emph{\bibinfo{journal}{JCAP}} \textbf{\bibinfo{volume}{04}},
  \bibinfo{pages}{012} (\bibinfo{year}{2019}).
\newblock \eprint{1809.09241}.

\bibitem{2019JCAP...06..047A}
\bibinfo{author}{{Armengaud}, E.} \emph{et~al.}
\newblock \bibinfo{title}{{Physics potential of the International Axion
  Observatory (IAXO)}}.
\newblock \emph{\bibinfo{journal}{JCAP}} \textbf{\bibinfo{volume}{2019}},
  \bibinfo{pages}{047} (\bibinfo{year}{2019}).
\newblock \eprint{1904.09155}.

\bibitem{1983PhLB..120..127P}
\bibinfo{author}{{Preskill}, J.}, \bibinfo{author}{{Wise}, M.~B.} \&
  \bibinfo{author}{{Wilczek}, F.}
\newblock \bibinfo{title}{{Cosmology of the invisible axion}}.
\newblock \emph{\bibinfo{journal}{Phys. Lett. B}}
  \textbf{\bibinfo{volume}{120}}, \bibinfo{pages}{127--132}
  (\bibinfo{year}{1983}).

\bibitem{1983PhLB..120..133A}
\bibinfo{author}{{Abbott}, L.~F.} \& \bibinfo{author}{{Sikivie}, P.}
\newblock \bibinfo{title}{{A cosmological bound on the invisible axion}}.
\newblock \emph{\bibinfo{journal}{Phys. Lett. B}}
  \textbf{\bibinfo{volume}{120}}, \bibinfo{pages}{133--136}
  (\bibinfo{year}{1983}).

\bibitem{1983PhLB..120..137D}
\bibinfo{author}{{Dine}, M.} \& \bibinfo{author}{{Fischler}, W.}
\newblock \bibinfo{title}{{The not-so-harmless axion}}.
\newblock \emph{\bibinfo{journal}{Phys. Lett. B}}
  \textbf{\bibinfo{volume}{120}}, \bibinfo{pages}{137--141}
  (\bibinfo{year}{1983}).

\bibitem{Kibble:1976sj}
\bibinfo{author}{Kibble, T. W.~B.}
\newblock \bibinfo{title}{{Topology of Cosmic Domains and Strings}}.
\newblock \emph{\bibinfo{journal}{J. Phys. A}} \textbf{\bibinfo{volume}{9}},
  \bibinfo{pages}{1387--1398} (\bibinfo{year}{1976}).

\bibitem{Zurek:1985qw}
\bibinfo{author}{Zurek, W.~H.}
\newblock \bibinfo{title}{{Cosmological Experiments in Superfluid Helium?}}
\newblock \emph{\bibinfo{journal}{Nature}} \textbf{\bibinfo{volume}{317}},
  \bibinfo{pages}{505--508} (\bibinfo{year}{1985}).

\bibitem{1988PhLB..205..228H}
\bibinfo{author}{{Hogan}, C.~J.} \& \bibinfo{author}{{Rees}, M.~J.}
\newblock \bibinfo{title}{{Axion miniclusters}}.
\newblock \emph{\bibinfo{journal}{Phys. Lett. B}}
  \textbf{\bibinfo{volume}{205}}, \bibinfo{pages}{228--230}
  (\bibinfo{year}{1988}).

\bibitem{khlopov_scalar}
\bibinfo{author}{Khlopov, M.}, \bibinfo{author}{Malomed, B.} \&
  \bibinfo{author}{Zeldovich, I.}
\newblock \bibinfo{title}{{Gravitational instability of scalar fields and
  formation of primordial black holes}}.
\newblock \emph{\bibinfo{journal}{Mon. Not. Roy. Astron. Soc.}}
  \textbf{\bibinfo{volume}{215}}, \bibinfo{pages}{575--589}
  (\bibinfo{year}{1985}).

\bibitem{Seidel:1991zh}
\bibinfo{author}{Seidel, E.} \& \bibinfo{author}{Suen, W.~M.}
\newblock \bibinfo{title}{{Oscillating soliton stars}}.
\newblock \emph{\bibinfo{journal}{Phys. Rev. Lett.}}
  \textbf{\bibinfo{volume}{66}}, \bibinfo{pages}{1659--1662}
  (\bibinfo{year}{1991}).

\bibitem{Guzman:2006yc}
\bibinfo{author}{Guzman, F.~S.} \& \bibinfo{author}{Urena-Lopez, L.~A.}
\newblock \bibinfo{title}{{Gravitational cooling of self-gravitating
  Bose-Condensates}}.
\newblock \emph{\bibinfo{journal}{Astrophys. J.}}
  \textbf{\bibinfo{volume}{645}}, \bibinfo{pages}{814--819}
  (\bibinfo{year}{2006}).
\newblock \eprint{astro-ph/0603613}.

\bibitem{2014NatPh..10..496S}
\bibinfo{author}{{Schive}, H.-Y.}, \bibinfo{author}{{Chiueh}, T.} \&
  \bibinfo{author}{{Broadhurst}, T.}
\newblock \bibinfo{title}{{Cosmic structure as the quantum interference of a
  coherent dark wave}}.
\newblock \emph{\bibinfo{journal}{Nature Physics}}
  \textbf{\bibinfo{volume}{10}}, \bibinfo{pages}{496--499}
  (\bibinfo{year}{2014}).
\newblock \eprint{1406.6586}.

\bibitem{Hui:2020hbq}
\bibinfo{author}{Hui, L.}, \bibinfo{author}{Joyce, A.},
  \bibinfo{author}{Landry, M.~J.} \& \bibinfo{author}{Li, X.}
\newblock \bibinfo{title}{{Vortices and waves in light dark matter}}.
\newblock \emph{\bibinfo{journal}{JCAP}} \textbf{\bibinfo{volume}{01}},
  \bibinfo{pages}{011} (\bibinfo{year}{2021}).
\newblock \eprint{2004.01188}.

\bibitem{PhysRevLett.107.233901}
\bibinfo{author}{Picozzi, A.} \& \bibinfo{author}{Garnier, J.}
\newblock \bibinfo{title}{Incoherent soliton turbulence in nonlocal nonlinear
  media}.
\newblock \emph{\bibinfo{journal}{Phys. Rev. Lett.}}
  \textbf{\bibinfo{volume}{107}}, \bibinfo{pages}{233901}
  (\bibinfo{year}{2011}).
\newblock
  \urlprefix\url{https://link.aps.org/doi/10.1103/PhysRevLett.107.233901}.

\bibitem{Mocz:2017wlg}
\bibinfo{author}{Mocz, P.} \emph{et~al.}
\newblock \bibinfo{title}{{Galaxy formation with BECDM \textendash{} I.
  Turbulence and relaxation of idealized haloes}}.
\newblock \emph{\bibinfo{journal}{Mon. Not. Roy. Astron. Soc.}}
  \textbf{\bibinfo{volume}{471}}, \bibinfo{pages}{4559--4570}
  (\bibinfo{year}{2017}).
\newblock \eprint{1705.05845}.

\bibitem{Mocz:2019pyf}
\bibinfo{author}{Mocz, P.} \emph{et~al.}
\newblock \bibinfo{title}{{First star-forming structures in fuzzy cosmic
  filaments}}.
\newblock \emph{\bibinfo{journal}{Phys. Rev. Lett.}}
  \textbf{\bibinfo{volume}{123}}, \bibinfo{pages}{141301}
  (\bibinfo{year}{2019}).
\newblock \eprint{1910.01653}.

\bibitem{Marsh:2018zyw}
\bibinfo{author}{Marsh, D. J.~E.} \& \bibinfo{author}{Niemeyer, J.~C.}
\newblock \bibinfo{title}{{Strong Constraints on Fuzzy Dark Matter from
  Ultrafaint Dwarf Galaxy Eridanus II}}.
\newblock \emph{\bibinfo{journal}{Phys. Rev. Lett.}}
  \textbf{\bibinfo{volume}{123}}, \bibinfo{pages}{051103}
  (\bibinfo{year}{2019}).
\newblock \eprint{1810.08543}.

\bibitem{Rogers:2020ltq}
\bibinfo{author}{Rogers, K.~K.} \& \bibinfo{author}{Peiris, H.~V.}
\newblock \bibinfo{title}{{Strong bound on canonical ultra-light axion dark
  matter from the Lyman-alpha forest}}.
\newblock \emph{\bibinfo{journal}{Phys. Rev. Lett.}}
  \textbf{\bibinfo{volume}{126}}, \bibinfo{pages}{071302}
  (\bibinfo{year}{2021}).
\newblock \eprint{2007.12705}.

\bibitem{Hlozek:2017zzf}
\bibinfo{author}{Hlozek, R.}, \bibinfo{author}{Marsh, D. J.~E.} \&
  \bibinfo{author}{Grin, D.}
\newblock \bibinfo{title}{{Using the Full Power of the Cosmic Microwave
  Background to Probe Axion Dark Matter}}.
\newblock \emph{\bibinfo{journal}{Mon. Not. Roy. Astron. Soc.}}
  \textbf{\bibinfo{volume}{476}}, \bibinfo{pages}{3063--3085}
  (\bibinfo{year}{2018}).
\newblock \eprint{1708.05681}.

\bibitem{Veltmaat:2018dfz}
\bibinfo{author}{Veltmaat, J.}, \bibinfo{author}{Niemeyer, J.~C.} \&
  \bibinfo{author}{Schwabe, B.}
\newblock \bibinfo{title}{{Formation and structure of ultralight bosonic dark
  matter halos}}.
\newblock \emph{\bibinfo{journal}{Phys. Rev. D}} \textbf{\bibinfo{volume}{98}},
  \bibinfo{pages}{043509} (\bibinfo{year}{2018}).
\newblock \eprint{1804.09647}.

\bibitem{chandra}
 \urlprefix\url{https://chandra.harvard.edu/photo/2017/sn1987a/}.

\bibitem{Penrose:1971uk}
\bibinfo{author}{Penrose, R.} \& \bibinfo{author}{Floyd, R.}
\newblock \bibinfo{title}{{Extraction of rotational energy from a black hole}}.
\newblock \emph{\bibinfo{journal}{Nature}} \textbf{\bibinfo{volume}{229}},
  \bibinfo{pages}{177--179} (\bibinfo{year}{1971}).

\bibitem{axiverse}
\bibinfo{author}{{Arvanitaki}, A.}, \bibinfo{author}{{Dimopoulos}, S.},
  \bibinfo{author}{{Dubovsky}, S.}, \bibinfo{author}{{Kaloper}, N.} \&
  \bibinfo{author}{{March-Russell}, J.}
\newblock \bibinfo{title}{{String axiverse}}.
\newblock \emph{\bibinfo{journal}{\prd}} \textbf{\bibinfo{volume}{81}},
  \bibinfo{pages}{123530} (\bibinfo{year}{2010}).
\newblock \eprint{0905.4720}.

\bibitem{Stott:2018opm}
\bibinfo{author}{Stott, M.~J.} \& \bibinfo{author}{Marsh, D. J.~E.}
\newblock \bibinfo{title}{{Black hole spin constraints on the mass spectrum and
  number of axionlike fields}}.
\newblock \emph{\bibinfo{journal}{Phys. Rev. D}} \textbf{\bibinfo{volume}{98}},
  \bibinfo{pages}{083006} (\bibinfo{year}{2018}).
\newblock \eprint{1805.02016}.

\bibitem{Arkani-Hamed:2006emk}
\bibinfo{author}{Arkani-Hamed, N.}, \bibinfo{author}{Motl, L.},
  \bibinfo{author}{Nicolis, A.} \& \bibinfo{author}{Vafa, C.}
\newblock \bibinfo{title}{{The String landscape, black holes and gravity as the
  weakest force}}.
\newblock \emph{\bibinfo{journal}{JHEP}} \textbf{\bibinfo{volume}{06}},
  \bibinfo{pages}{060} (\bibinfo{year}{2007}).
\newblock \eprint{hep-th/0601001}.

\bibitem{Raffelt:2006cw}
\bibinfo{author}{Raffelt, G.~G.}
\newblock \bibinfo{title}{{Astrophysical axion bounds}}.
\newblock \emph{\bibinfo{journal}{Lect. Notes Phys.}}
  \textbf{\bibinfo{volume}{741}}, \bibinfo{pages}{51--71}
  (\bibinfo{year}{2008}).
\newblock \eprint{hep-ph/0611350}.

\bibitem{Klaer:2017ond}
\bibinfo{author}{Klaer, V.~B.} \& \bibinfo{author}{Moore, G.~D.}
\newblock \bibinfo{title}{{The dark-matter axion mass}}.
\newblock \emph{\bibinfo{journal}{JCAP}} \textbf{\bibinfo{volume}{11}},
  \bibinfo{pages}{049} (\bibinfo{year}{2017}).
\newblock \eprint{1708.07521}.

\bibitem{Gorghetto:2020qws}
\bibinfo{author}{Gorghetto, M.}, \bibinfo{author}{Hardy, E.} \&
  \bibinfo{author}{Villadoro, G.}
\newblock \bibinfo{title}{{More Axions from Strings}}.
\newblock \emph{\bibinfo{journal}{SciPost Phys.}}
  \textbf{\bibinfo{volume}{10}}, \bibinfo{pages}{050} (\bibinfo{year}{2021}).
\newblock \eprint{2007.04990}.

\bibitem{Hiramatsu:2012sc}
\bibinfo{author}{Hiramatsu, T.}, \bibinfo{author}{Kawasaki, M.},
  \bibinfo{author}{Saikawa, K.} \& \bibinfo{author}{Sekiguchi, T.}
\newblock \bibinfo{title}{{Axion cosmology with long-lived domain walls}}.
\newblock \emph{\bibinfo{journal}{JCAP}} \textbf{\bibinfo{volume}{01}},
  \bibinfo{pages}{001} (\bibinfo{year}{2013}).
\newblock \eprint{1207.3166}.

\bibitem{Svrcek:2006yi}
\bibinfo{author}{Svrcek, P.} \& \bibinfo{author}{Witten, E.}
\newblock \bibinfo{title}{{Axions In String Theory}}.
\newblock \emph{\bibinfo{journal}{JHEP}} \textbf{\bibinfo{volume}{06}},
  \bibinfo{pages}{051} (\bibinfo{year}{2006}).
\newblock \eprint{hep-th/0605206}.

\bibitem{PhysRevLett.58.1799}
\bibinfo{author}{Wilczek, F.}
\newblock \bibinfo{title}{Two applications of axion electrodynamics}.
\newblock \emph{\bibinfo{journal}{Phys. Rev. Lett.}}
  \textbf{\bibinfo{volume}{58}}, \bibinfo{pages}{1799--1802}
  (\bibinfo{year}{1987}).
\newblock \urlprefix\url{https://link.aps.org/doi/10.1103/PhysRevLett.58.1799}.

\bibitem{2010NatPh...6..284L}
\bibinfo{author}{{Li}, R.}, \bibinfo{author}{{Wang}, J.},
  \bibinfo{author}{{Qi}, X.-L.} \& \bibinfo{author}{{Zhang}, S.-C.}
\newblock \bibinfo{title}{{Dynamical axion field in topological magnetic
  insulators}}.
\newblock \emph{\bibinfo{journal}{Nature Physics}}
  \textbf{\bibinfo{volume}{6}}, \bibinfo{pages}{284--288}
  (\bibinfo{year}{2010}).
\newblock \eprint{0908.1537}.

\bibitem{Zhang_2020}
\bibinfo{author}{Zhang, J.} \emph{et~al.}
\newblock \bibinfo{title}{{Large Dynamical Axion Field in Topological
  Antiferromagnetic Insulator Mn$_2$Bi$_2$Te$_5$}}.
\newblock \emph{\bibinfo{journal}{Chinese Physics Letters}}
  \textbf{\bibinfo{volume}{37}}, \bibinfo{pages}{077304}
  (\bibinfo{year}{2020}).
\newblock
  \urlprefix\url{https://doi.org/10.1088%2F0256-307x%2F37%2F7%2F077304}.

\bibitem{2018PhLA..382.3018C}
\bibinfo{author}{{Cheng}, Y.}, \bibinfo{author}{{Peng}, B.},
  \bibinfo{author}{{Hu}, Z.}, \bibinfo{author}{{Zhou}, Z.} \&
  \bibinfo{author}{{Liu}, M.}
\newblock \bibinfo{title}{{Recent development and status of magnetoelectric
  materials and devices}}.
\newblock \emph{\bibinfo{journal}{Physics Letters A}}
  \textbf{\bibinfo{volume}{382}}, \bibinfo{pages}{3018--3025}
  (\bibinfo{year}{2018}).

\bibitem{OHare:2017yze}
\bibinfo{author}{O'Hare, C. A.~J.} \& \bibinfo{author}{Green, A.~M.}
\newblock \bibinfo{title}{{Axion astronomy with microwave cavity experiments}}.
\newblock \emph{\bibinfo{journal}{Phys. Rev.}} \textbf{\bibinfo{volume}{D95}},
  \bibinfo{pages}{063017} (\bibinfo{year}{2017}).
\newblock \eprint{1701.03118}.

\bibitem{Frank:2016qwj}
\bibinfo{author}{Frank, K.~A.} \emph{et~al.}
\newblock \bibinfo{title}{{Chandra Observes the end of an era in SN 1987a}}.
\newblock \emph{\bibinfo{journal}{Astrophys. J.}}
  \textbf{\bibinfo{volume}{829}}, \bibinfo{pages}{40} (\bibinfo{year}{2016}).
\newblock \eprint{1608.02160}.

\bibitem{Orlando:2015dfa}
\bibinfo{author}{Orlando, S.}, \bibinfo{author}{Miceli, M.},
  \bibinfo{author}{Pumo, M.~L.} \& \bibinfo{author}{Bocchino, F.}
\newblock \bibinfo{title}{{Supernova 1987A: a Template to Link Supernovae to
  their Remnants}}.
\newblock \emph{\bibinfo{journal}{Astrophys. J.}}
  \textbf{\bibinfo{volume}{810}}, \bibinfo{pages}{168} (\bibinfo{year}{2015}).
\newblock \eprint{1508.02275}.

\bibitem{Indebetouw:2013sva}
\bibinfo{author}{Indebetouw, R.} \emph{et~al.}
\newblock \bibinfo{title}{{Dust Production and Particle Acceleration in
  Supernova 1987A Revealed with ALMA}}.
\newblock \emph{\bibinfo{journal}{Astrophys. J. Lett.}}
  \textbf{\bibinfo{volume}{782}}, \bibinfo{pages}{L2} (\bibinfo{year}{2014}).
\newblock \eprint{1312.4086}.

\bibitem{PhysRevX.9.041057}
\bibinfo{author}{Dematteis, G.}, \bibinfo{author}{Grafke, T.},
  \bibinfo{author}{Onorato, M.} \& \bibinfo{author}{Vanden-Eijnden, E.}
\newblock \bibinfo{title}{Experimental evidence of hydrodynamic instantons: The
  universal route to rogue waves}.
\newblock \emph{\bibinfo{journal}{Phys. Rev. X}} \textbf{\bibinfo{volume}{9}},
  \bibinfo{pages}{041057} (\bibinfo{year}{2019}).
\newblock \urlprefix\url{https://link.aps.org/doi/10.1103/PhysRevX.9.041057}.

\bibitem{1988assy.book.....C}
\bibinfo{author}{{Coleman}, S.}
\newblock \emph{\bibinfo{title}{{Aspects of Symmetry}}}
  (\bibinfo{publisher}{Cambridge University Press}, \bibinfo{year}{1988}).

\bibitem{Vafa:1984xg}
\bibinfo{author}{Vafa, C.} \& \bibinfo{author}{Witten, E.}
\newblock \bibinfo{title}{{Parity Conservation in QCD}}.
\newblock \emph{\bibinfo{journal}{Phys. Rev. Lett.}}
  \textbf{\bibinfo{volume}{53}}, \bibinfo{pages}{535} (\bibinfo{year}{1984}).

\bibitem{GEORGI1986241}
\bibinfo{author}{Georgi, H.} \& \bibinfo{author}{Randall, L.}
\newblock \bibinfo{title}{Flavor conserving cp violation in invisible axion
  models}.
\newblock \emph{\bibinfo{journal}{Nuclear Physics B}}
  \textbf{\bibinfo{volume}{276}}, \bibinfo{pages}{241--252}
  (\bibinfo{year}{1986}).
\newblock
  \urlprefix\url{https://www.sciencedirect.com/science/article/pii/0550321386900222}.

\end{thebibliography}

\newpage
\appendix

\beginsupplement
\section{Supplemental: Theoretical Methods}

\subsection{The Axion and Quantum Chromodynamics}

We look here in more detail at the strong $CP$ problem and the axion. When we write down the Lagrangian density for Quantum Chromodynamics (QCD) we must include all terms allowed by the symmetries of the theory. This means that the Lagrangian density should include the $CP$-violating term {In the supplemental material, we use natural units with $c = \hslash = 1$.}

\begin{equation}
    \mathcal{L} \supset \bar{\theta} \frac{g_3^2}{32 \pi^2} G_{\mu \nu} \cdot \tilde{G}^{\mu \nu},
    \label{CPV}
\end{equation}
where $g_3$ is the coupling constant of the strong force, $G_{\mu \nu}$ is the gluon field strength tensor, $\tilde{G}_{\mu \nu} = \epsilon_{\mu \nu \rho \sigma} G^{\rho \sigma}$ is its dual, and $\bar{\theta}$ includes contributions from the quark masses and from the structure of the vacuum itself:

\begin{equation}
    \Bar{\theta} = \theta + {\rm arg} \, {\rm det} (M),
\end{equation}
where $\theta$ is the QCD vacuum angle and $M$ is the quark mass matrix. We seek an explanation for why $\bar{\theta}$ is so small.\\

To this end, we introduce the global Peccei-Quinn symmetry $U(1)_{PQ}$. {The spontaneous breaking of this symmetry yields} a massless Nambu-Goldstone boson, the axion. This means that under a $U(1)_{PQ}$ transformation with parameter $\epsilon$ the axion field transforms like:

\begin{equation}
    a(x) \xrightarrow{} a(x) + \epsilon f_a.
\end{equation}
When the $U(1)_{PQ}$ symmetry is exact, the axion must be exactly massless, so that adding $\epsilon f_a$ to the axion field results in the same energy density. 

The axion interacts with gluons via the Lagrangian density
\begin{equation}
\mathcal{L} \supset \frac{a}{f_a} \frac{g_3^2}{32 \pi^2} G_{\mu \nu} \cdot \tilde{G}^{\mu \nu}.
\label{eqn:axion_gluon}
\end{equation}
Notice that this has exactly the same form as the problematic strong $CP$-violating term (\ref{CPV}), but the coefficient of $G \cdot \tilde{G}$ is now a dynamical field. This is what allows the total coefficient of $G \cdot \tilde{G}$ to relax to zero in the ground state of the theory. How this term is generated is discussed in more detail in the next subsection. We note in passing that the axion's coupling to photons takes the same form $\mathcal{L} \supset \frac{g_{a \gamma \gamma}}{4} F \cdot \tilde{F} = g_{a \gamma \gamma} {\bf E} \cdot {\bf B}$, where $F$ is the electromagnetic field strength tensor.\\

The axion's coupling to gluons also ensures that the $U(1)_{PQ}$ symmetry is {\it anomalous}. An anomalous symmetry is a symmetry that would be conserved in a classical theory but is broken by quantum effects. Put another way, the theory's Lagrangian is invariant under the symmetry, but the path inegral measure is not. This means that the axion is no longer massless, but is a pseudo-Nambu-Goldstone boson. The potential energy, and therefore the mass, of the axion field is generated by QCD instantons. Instantons are non-trivial solutions to the classical equations of motion, whose effects are not captured by perturbative quantum field theory methods. Instantons have been observed experimentally in the context of fluid dynamics \cite{PhysRevX.9.041057}.

{The effects of instantons in QCD} can be accounted for by a $\theta$-dependent change to the vacuum energy, leading to an effective potential for the axion. In the dilute instanton gas approximation (see e.g. Ref.~\cite{1988assy.book.....C}) applicable to high-temperature QCD, the potential is:
\begin{equation}
    V(a(x))\propto -{\rm cos} \left( \bar{\theta} + \frac{a(x)}{f_a} \right).
\end{equation}
Therefore, {minimising the potential energy of} the axion field sets the total coefficient of $G \cdot \tilde{G}$ to zero. The mass of the axion is found from a Taylor expansion of $V(a(x))$. The constant of proportionality is the QCD topological susceptibility, which is temperature-dependent and can be computed by instanton methods, in {effective field theory}, or using non-perturbative lattice QCD.

How do we know that the minimum of the axion potential is at precisely the value that leads to no $CP$ violation in strong interactions? One way to see this is via the Vafa-Witten theorem, which tells us that in vector-like theories, such as QCD, dynamical parity-violating terms are zero in the ground state of the theory \cite{Vafa:1984xg}. The neutron EDM violates $P$ as well as $CP$, and the axion field make the total neutron EDM a dynamical variable. Therefore, {\it in the absence of weak interactions}, the Vafa-Witten theorem guarantees that the axion field potential minimum will be at zero neutron EDM. However, the theory of quarks is not quite vector-like because the quarks are also charged under the chiral weak interaction. This means that the minimum of the axion potential is slightly shifted from the $CP$ conserving value. The magnitude of this effect can be calculated using effective field theory, and is found to lead to an effective $CP$ violating angle $\bar{\theta} \simeq \mathcal{O}(10^{-17})$~\cite{GEORGI1986241}, far below current experimental bounds. 

\subsection{Invisible axion models}
\label{sec:InvisibleAxionModels}

\begin{figure}
    \centering
    \includegraphics[width=0.5\textwidth]{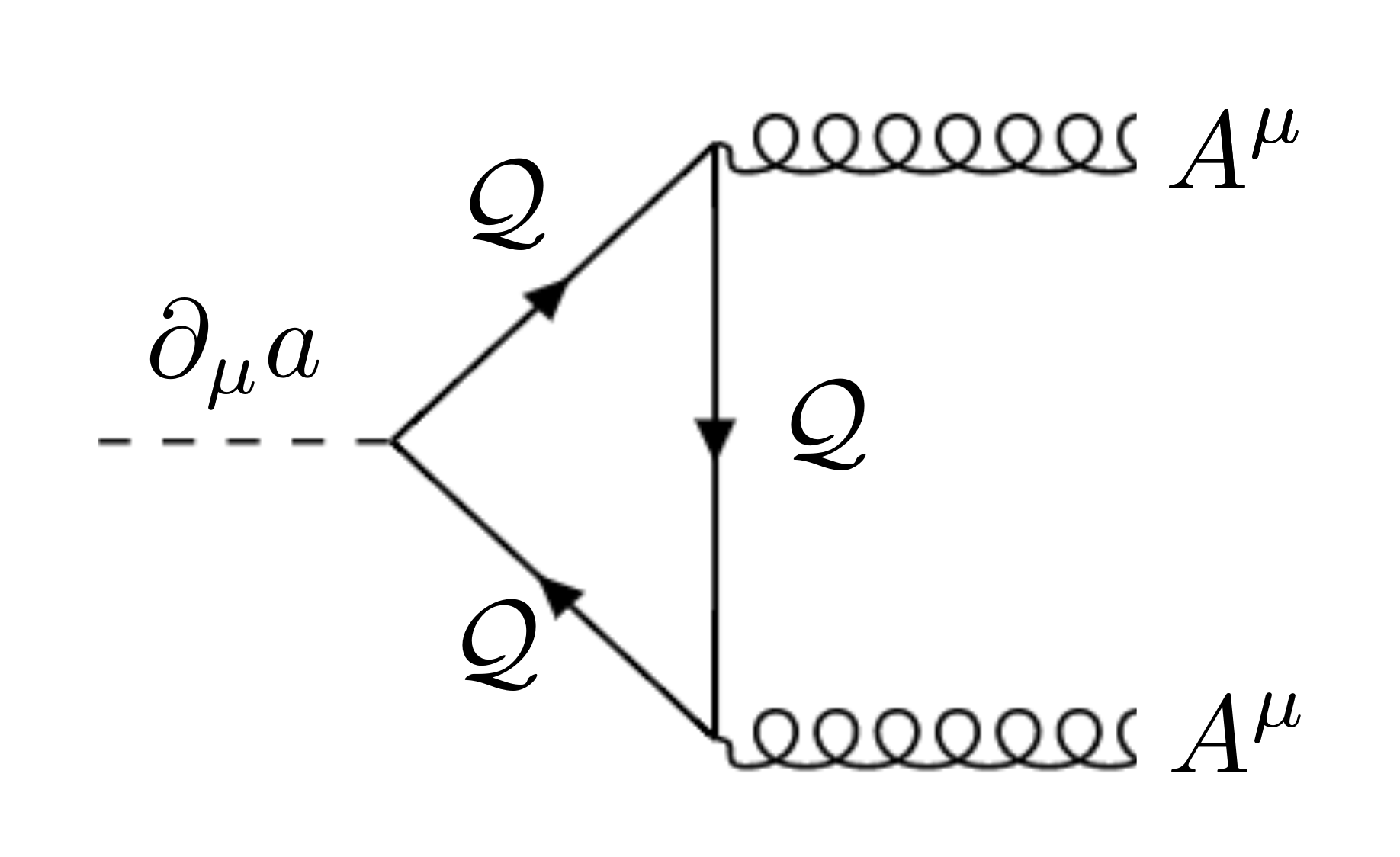}
    \caption{The colour anomaly of the Peccei-Quinn symmetry. The current of the PQ symmetry is $J_\mu^{\rm PQ}=\partial_\mu a$. Due to the chiral interaction between the axion and a quark, $\mathcal{Q}$ (under which the left- and right-handed Weyl spinors carry opposite charge), this current is not conserved in the quantum theory. The loop diagram mediates a non-vanishing interaction between axions and gluons, $A^\mu$.} 
    \label{fig:colour_anomaly}
\end{figure}

So far, we have described in only general terms how the axion solves the strong $CP$ problem. Now, we would like to be more specific. The key property of the axion is its coupling to gluons described above. In order to solve the strong $CP$ problem, this coupling should be generated by the {\em chiral anomaly} mediated by the coupling between the axion and fermions charged under the strong interactions, leading to the triangle Feynman diagram in Fig.~\ref{fig:colour_anomaly}. In the absence of the chiral anomaly, this diagram would have zero amplitude. In the Standard Model, the fermions charged undeer the strong interactions are the quarks, $u,d,c,s,t,b$. 

In the original axion model \cite{weinberg1978,wilczek1978}, the Peccei-Quinn field $\Phi$ is a second Higgs field, with one field giving mass to up-type quarks, and another to down-type quarks. The axion field is then the complex phase of the combined Higgs field. However, given the known vacuum expectation value of the Higgs field, this model is ruled out by experimental data from colliders. This led to the development of ``invisible'' axion models. In this section we will describe the first two invisible axion models {A review of these and more general models can be found in Ref.~\cite{DiLuzio:2020wdo}.}. In these models, the Peccei-Quinn field is not related to the Higgs field, and so its vacuum expectation value $v_{\rm PQ}$ is a free parameter. If $v_{\rm PQ}$ is high enough, the axion's interactions are weak enough to have evaded experiments to date {The axion decay constant is given by $f_a=v_{\rm PQ}/N$, where $N$ is the (model dependent) color anomaly of the Peccei-Quinn symmetry}.\\

We describe first the Kim~\cite{PhysRevLett.43.103} Shifman-Vainshtein-Zakharov~\cite{SHIFMAN1980493} (KSVZ) model. The KSVZ model extends the Standard Model by the introduction of two new fields. The first is the Peccei-Quinn field $\Phi$, which is uncharged under all the Standard Model interactions. The second is a new heavy vector fermion $\mathcal{Q}$, which is charged under the strong interactions, i.e., a new exotic quark.{The historical KSVZ model also has an electrical charge for the new quark, with a value inspired by grand unified theories. For simplicity, we do not discuss further this option.} The Lagrangian of this sector is
\begin{equation}
    \mathcal{L} = (\partial_\mu\Phi)(\partial^\mu \Phi)-\frac{\lambda}{4}(|\Phi|^2-v_{\rm PQ}^2)^2 + i\bar{\mathcal{Q}}(\partial_\mu - ig_3 A_\mu) \gamma^\mu \mathcal{Q}+(y_Q\bar{\mathcal{Q}}_L\mathcal{Q}_R\Phi+y_Q^\dagger\bar{\mathcal{Q}}_R\mathcal{Q}_L\Phi^\dagger)\, .
\end{equation}
The first term is the kinetic term of the PQ field. The second term is the potential of the PQ field, with the characteristic symmetry breaking form, with $\lambda$ the PQ self-interaction strength. In the vacuum, the field takes the form $\Phi = (v_{\rm PQ}+\chi)e^{ia/v_{\rm PQ}}$, with $\chi$ a massive field and $a$ the massless Nambu-Goldstone boson arising from breaking of the rotational symmetry in the complex plane of $\Phi$, i.e., $U(1)$ symmetry. The third term is the quark kinetic term, with the $SU(3)$ covariant derivative: $A_\mu$ is the matrix gluon field, an element of $SU(3)$ (which appears in the right-hand side of the Feynman diagram). Note that the new quark does not couple to the Higgs boson, and so has no mass term in the Lagrangian. 

The fourth term is where the real action of the PQ mechanism resides: the coupling between the PQ field and the new quark, with a Yukawa coupling constant $y_Q$. This term gives a mass to the new quark when $\Phi$ gains a vacuum expectation value: $\Phi$ acts like the Higgs for the new quark. If the product $y_Q v_{\rm PQ}$ is large, then the new quark is too heavy to have been observed in any collider experiment so far. This interaction term also leads to a coupling between the axion and the new heavy quark, leading to the required chiral anomaly (this gives the left-hand side of the Feynman diagram). \\

The Dine-Fischler-Srednicki~\cite{Dine:1981rt} Zhitnitsky~\cite{Zhitnitsky:1980tq} (DFSZ) model is slightly more subtle to describe, since it involves changes to the Higgs sector of the Standard Model, splitting the Higgs into up- and down-type doublets. This additional subtlety avoids the introduction of the new quark, $\mathcal{Q}$, of the KSVZ model, and instead mediates the coupling between axions and gluons using the quarks of the Standard Model. We do not discuss this model in detail, since it requires detailed background on the Higgs and the Standard Model. From the perspective of low-energy phenomenology, the DFSZ model differs from the KSVZ model in that it possesses a direct coupling between the axion and the electron, $g_{aee}$, {whereas the KSVZ model acquires this coupling only at loop order, and so it is suppressed in magnitude}. Since the electron and the quarks of the Standard Model also carry electric charge, the DFSZ axion acquires its coupling to electromagnetism via loop diagrams mediated by the Standard Model fermions, in addition to its mixing with the neutral pion, thus leading to a different strength of coupling, $g_{a\gamma\gamma}$ than that in the KSVZ model (see below). These two differences, in the values of $g_{aee}$ and $g_{a\gamma\gamma}$ , will allow (in the event that the axion is discovered in the future) for the KSVZ and DFSZ models (and other axion variants) to be distinguished by measurements as described in the experimental companion to this review.\\

Many axion search strategies depend on the axion's mass and coupling to electromagnetism. These are related by \cite{diCortona:2015ldu}
\begin{equation}
    g_{a \gamma \gamma} = \frac{\alpha_{em}}{2 \pi f_a} \left(\frac{\mathcal{E}}{N} - 1.92(4) \right) = \left(0.203(3) \frac{\mathcal{E}}{N} - 0.39(1) \right) \frac{m_a}{{\rm GeV}^2},
    \label{eq:EMcoupling}
\end{equation}
where the numbers in brackets denote theoretical uncertainties, $\alpha_{em}$ is the electromagnetic fine structure constant, $\mathcal{E}$ is the electromagnetic anomaly of the PQ symmetry and $N$ the colour anomaly. The model dependent coupling coefficient $ C_{\gamma} = \frac{\mathcal{E}}{N} - 1.92(4)$ is often used. The first term in Eq.~(\ref{eq:EMcoupling}) is generated by the mixing between the axion and the neutral pion, and is model-independent (so long as the axion has a coupling to gluons and solves the strong $CP$ problem: the coupling can be absent for an ALP), while $\mathcal{E}$ and $N$ in the second term are model-dependent. For example, in the DFSZ model $\frac{\mathcal{E}}{N} = \frac{8}{3}$ (generated by the fermions charged under both electromagnetism and PQ), whereas for the KSVZ model described here $\frac{\mathcal{E}}{N} = 0$. In principle, $\frac{\mathcal{E}}{N}$ could be large, leading to a large value for $|g_{a \gamma \gamma}|$ (there is some maximum set by unitarity). Alternatively, $\frac{\mathcal{E}}{N}$ could by chance be such that the two terms in Eq.~\eqref{eq:EMcoupling} approximately cancel, leading to a very small value for $|g_{a \gamma \gamma}|$. However, such a model would need to be deliberately constructed so as to hide the axion, and in general we take the DFSZ model to give the smallest canonical value of $|g_{a \gamma \gamma}|$. Precisely outlining the allowed range of $g_{a \gamma \gamma}$ in viable models is an active area of research.

\subsection{Equations of Axion Dark Matter}

The Friedmann equation: 
\begin{equation}
    \frac{1}{R(t)^2}\left(\frac{dR}{dt}\right)^2\equiv H(t)^2 = \frac{8\pi^3 G_N}{90}g_\star (T)T^4\, ,
    \label{eqn:friedmann}
\end{equation}
where $t$ is time ``since the Big Bang'', $T$ is temperature, $G_N$ is the Newton constant, $g_\star(T)$ counts the number of relativistic particles at temperature $T$ {(with different factors for bosons and fermions)}, and we have defined the Hubble rate $H(t)$ from the cosmic scale factor $R(t)$. The evolution of temperature with time, and also $g_\star(T)$, can be found from elementary thermodynamic considerations. {At the beginning of the hot Big Bang, the temperture was at its maximum value, $T_\textup{hot}$, the time was $t_\textup{hot}$, and the value of $R(t_\textup{hot})$ is chosen by normalisation.}

In the simple case described by Scenario A, the axion field evovles according to the Klein-Gordon equation:
\begin{equation}
    \frac{d^2a}{dt^2}+3H(t)\frac{da}{dt}+m_a(T)^2a = 0\, ,
\end{equation}
with initial condition $\theta(t_{\textup{ hot}})=\theta_i$, $\dot{\theta}(t_{\textup{ hot}})=0$, and the axion number density is
\begin{equation}
    n_a = \frac{1}{m_a(T)}\left[\frac{1}{2}\left(\frac{da}{dt}\right)^2+\frac{1}{2}m_a(T)^2a^2\right]\, .
\end{equation}
The temperature dependence of the axion mass, $m_a(T)$, {can be computed from the topological susceptibility as described briefly above.} {At temperatures above the QCD phase transition the mass grows with temperature, roughly described by a power law, $m_a\propto T^4$, while at low temperature it has the asymptotic} value given in Eq.~(\ref{eq: mass}). The Hubble parameter acts as a friction term in the Klein-Gordon equation, which falls over time as the Universe expands and cools. The axion field begins damped harmonic motion at the epoch when $H(T)\approx m_a(T)$ {(labelled $t_\textup{dyn}$ in Fig.~\ref{fig:two_roads})}. At some slightly later time, $t_{\rm cold}$, the oscillation period of the axion field becomes large compared to the Hubble rate, and the Klein-Gordon equation enters an attractor solution in which the number density is conserved modulo the expansion of the Universe, i.e., $n_a(t)/R(t)^3=\text{ const.}$. The axion dark matter density is then given by $\rho_a=m_a n_a$, and is a function of the zero-temperature axion mass and the initial field value.  

At late times the axion is described by the Gross-Piatevski-Poisson equations. These equations arise as the non-relativistic limit of the Klein-Gordon equation for the axion field in a weakly perturbed Friedmann-Robertson-Walker spacetime, under a phase approximation, $a=\sqrt{2/m_a}(\psi e^{im_at/\hslash}+\psi^* e^{-im_at/\hslash})$. The ``wave function'' $\psi$ obeys: 
\begin{align}
    i\hslash\frac{\partial\psi}{\partial t}+\frac{\hslash^2}{2 m_a}\nabla^2 \psi -\frac{\lambda_a}{2m_a} |\psi|^2\psi -m V_N\psi = 0 \, , \\
    \nabla^2 V_N = 4\pi G_N (m_a |\psi|^2 + \rho)\, , 
\end{align}
where $\lambda_a$ is the dimensionless axion self-interaction coupling constant. In the dilute instanton gas approximation, $\lambda_a=-m_a^2/f_a^2$. The mass density of axions to leading order is $m_a|\psi|^2$. The Newtonian gravitational potential is $V_N$, and $\rho$ represents the density of all other gravitating matter.

\end{document}